\documentclass[usenatbib,usegraphicx]{mn2e}
\usepackage{bm}
\usepackage{morefloats}
\usepackage{hyperref}

\title[$N$-body models of GCs: Masses, M/L ratios and IMBHs]
{$N$-body modeling of globular clusters: Masses, mass-to-light ratios and intermediate-mass black holes}
\author[Baumgardt]{H. Baumgardt\thanks{E-mail:
h.baumgardt@uq.edu.au}\\
School of Mathematics and Physics, The University of Queensland, St. Lucia, QLD 4072, Australia \\
}

\begin{document}

\date{Accepted 2016 xx xx. Received 2016 xx xx; in original form 2016 xx xx}

\pagerange{\pageref{firstpage}--\pageref{lastpage}} \pubyear{201x}

\maketitle

\label{firstpage}

\begin{abstract}
We have determined the masses and mass-to-light ratios of 50 Galactic globular clusters by comparing their velocity dispersion and surface brightness
profiles against a large grid of 900 $N$-body simulations of star clusters of varying initial concentration, size and central black
hole mass fraction. Our models follow the evolution of the clusters under the combined effects of
stellar evolution and two-body relaxation allowing us to take the effects of mass segregation and energy equipartition 
between stars self-consistently into account. For a subset of 16 well observed clusters we also derive their kinematic distances. 
We find an average mass-to-light ratio
of Galactic globular clusters of $<M/L_V>=1.98 \pm 0.03$, which agrees very well with the expected $M/L$ ratio if the initial mass function
(IMF) of the clusters was a standard Kroupa or Chabrier mass function.
We do not find evidence for a decrease of the average mass-to-light ratio with metallicity.
The surface brightness and
velocity dispersion profiles of most globular clusters are incompatible with the presence of intermediate-mass black holes (IMBHs) with
more than a few thousand $M_\odot$ in them.
The only clear exception is $\omega$~Cen, where the velocity dispersion profile provides strong evidence for the presence of a $\sim$40,000 M$_\odot$~IMBH
in the centre of the cluster. 
\end{abstract}


\section{Introduction} \label{sec:intro}

Globular clusters are among the oldest structures in the universe, having formed within 1 to 2 Gyr after the Big Bang 
\citep{kravtsovgnedin2005}. Studying their origin and evolution has therefore important implications for our understanding
of star formation and the growth of structure in the early universe. In addition, due to their high central densities and high stellar encounter rates, 
globular clusters are also unique environments for the creation of exotic stars like blue stragglers \citep{bailyn1995,davies2004}, 
low-mass X-ray binaries \citep{verbunt1993,pooleyetal2003} and millisecond pulsars \citep{manchesteretal1991}. The high stellar 
densities in globular cluster could also give rise to the creation of intermediate-mass 
black holes \citep{pzm2002,pzetal2004,gierszetal2015}, which might be the progenitors of supermassive black holes in Galactic centers.
Globular clusters are finally important environments for the 
creation of tight black hole binaries which merge through the emission of gravitational waves \citep{banerjeeetal2010,downingetal2011,
rodriguezetal2016a,rodriguezetal2016b,askaretal2016}.

In order to understand the rate of creation of exotic stars, it is important to know the mass density profile of globular clusters
and how different types of stars are distributed within a globular cluster. This is possible by a detailed modeling of the internal 
kinematics of globular clusters. Several
methods have been suggested in the literature to derive cluster masses from observed density profiles: One can either using analytic formulas
which relate a cluster's mass to its radius and velocity dispersion inside some radius \citep[e.g.][]{mandushevetal1991,straderetal2011},
or fit analytic density profiles like Plummer or King models to the observed velocity and surface density profiles
of globular clusters \citep[e.g.][]{mclaughlinvandermarel2005,kimmigetal2015}. Finally it is possible to deproject the observed
surface density profile and then derive the cluster mass through Jeans modeling and a fit of the observed velocity dispersion profile
\citep[e.g.][]{vandevenetal2006,noyolaetal2008,lutzgendorfetal2012,lutzgendorfetal2013}.

Most approaches assume a constant mass-to-light ratio inside globular clusters. However, since the relaxation times of globular 
clusters are generally much smaller 
than their ages, high-mass stars like compact remnants and giant stars are concentrated towards the cluster centers while low-mass
stars are pushed towards the outer cluster parts \citep{baumgardtmakino2003}. Hence the assumption of a constant mass-to-light
ratio is not valid for globular clusters. In addition, due to energy equipartition, massive stars move more slowly at a given radius 
compared to average cluster stars \citep{trentivandermarel2013,bianchinietal2016}. As a result, the velocity dispersion derived from giant stars 
will underestimate the true velocity dispersion, which leads to an underestimation of the total cluster mass if mass segregation is
not properly taken into account \citep{shanahangieles2015}.

It is possible to account for mass segregation by e.g. using multi-mass King-Michie models \citep{michie1963,gg1979} or the more recently
suggested {\tt LIMEPY} models \citep{gieleszocchi2015,zocchietal2016b}. Multi-mass models have however additional degrees of freedom 
since the amount of mass segregation between different mass components can in principle be freely chosen in the models.

In the present paper we follow a different approach to 
derive the absolute masses and mass-to-light ratios of globular clusters from their surface density and velocity dispersion profiles. We 
perform a large grid of $N$-body simulations and scale each model so that it has the same half-light radius as the observed clusters. Scaling is done in
such a way that the relaxation time is kept constant, thereby making sure that mass segregation of stars and (partial) energy equipartition 
between them are taken into account in a self consistent way in the scaled models, i.e. each model has the exact amount of mass segregation
which a real globular cluster would have if it started from the same initial condition. We then determine the model which best fits the observed density and velocity
dispersion profile for each globular cluster and determine the total mass, mass-to-light ratio and the possible presence of an intermediate mass
black hole in the observed clusters from the best-fitting model. Our paper is organised as follows:
In section 2 we describe the grid of $N$-body models that we have performed, and in section 3 we describe the selection of the observational data.
Section 4 presents our results and we draw our conclusions in section~5. 

\section{The $N$-body models}

In total we calculated a grid of $\sim$900 $N$-body simulations, varying the initial density profile, half-mass radius $r_h$, cluster metallicity [Fe/H] and 
the mass fraction $M_{BH}/M_{GC}$ of a central IMBH between the different simulations. Our clusters did not contain primordial binaries, however binaries could form
dynamically during the simulations. 
All simulations were made using the GPU-enabled version of the collisional $N$-body code NBODY6 \citep{aarseth1999,nitadoriaarseth2012}.
Clusters without IMBHs and clusters with 
IMBH mass fractions of $M_{BH}/M_{GC}=0.01$ and $M_{BH}/M_{GC}=0.02$
started with $N=100,000$ stars, while clusters with an IMBH mass fraction of $M_{BH}/M_{GC}=0.005$ were run with $N=200,000$ stars initially.
In total we performed 720 simulations with $N=100,000$ stars and 48 simulations with $N=200,000$ stars.
We also performed test simulations with $N=50,000$ stars to test the dependency of our results on the initial number of cluster stars, but found that
the initial particle number has a negligible influence on the results. 

The initial density profiles of our clusters were given by \citet{king1962} models with initial dimensionless central concentrations 
$c=\log{r_c/r_t}$ of $c =$0.2, 0.5, 1.0, 1.5, 2.0 and 2.5
respectively.  We also simulated clusters starting with \citet{king1966} density profiles, but found that these led to clusters with a too small variation
in the final density profile which cannot fit observed surface density profiles for a significant fraction of globular clusters. 
Initial cluster models were set up using the method described in \citet{hilkeretal2007}, by first deprojecting the density profile, then calculating the 
distribution function $f(E)$ and finally choosing particle positions and velocities.
We used 8 grid points for the initial half-mass radius $r_h$
given by $r_h=$~2, 3, 5, 7, 10, 15, 25 and 35 pc for the $N=50,000$ star clusters. For the $N=100,000$ and $N=200,000$ star clusters, the initial
half-mass radii were reduced by
factors of 0.836 and 0.696 respectively so that these clusters have the same initial relaxation time than the corresponding $N=50,000$ star models.
For each value of $r_h$ and $c$ we ran three simulations starting from different random number seeds to increase the statistical significance of our results.

Stellar evolution was modeled according to the stellar evolution routines of \citet{hurleyetal2000}, assuming black hole and neutron star retention fractions of 10\%.
All clusters started with stars distributed according to a \citet{kroupa2001} mass function with lower and upper mass limits of 0.1~$M_\odot$ and 100 M$_\odot$
respectively.
For clusters without IMBHs, we ran simulations at three different metallicities given by [Fe/H]=-1.8, -1.3 and -0.7 respectively. For the later comparison
with observed clusters we always use those clusters from our grid that are closest in metallicity to the metallicity of the observed clusters. This should be accurate
enough since metallicity dependent effects on the internal cluster evolution are largely removed due to our scaling procedure described below so that the
influence of cluster metallicity on our results (e.g. cluster mass) is small.

Simulations were run up to an age of $T=13.5$ Gyrs, and we stored data spaced by $T=50$ Myrs for all times between $T=10.5$ and $T=13.5$ Gyr. In order to compare our
grid of simulations to observed clusters, we combined 10 snapshots spanning a $T=500$ Myr time span centered around the age of each cluster. Since we
ran three different realizations for each grid point, our final models after combining the individual snapshots contained roughly $3 \cdot 10^6$ stars per grid point,
which is larger than the actual number of stars in most observed clusters.

For [Fe/H]=-1.3 we also ran simulations with central IMBHs, choosing the IMBH masses such that the mass ratio of the IMBH to the total cluster
mass at the end of our simulations ($T=13.5$ Gyr) was equal to $M_{IMBH}/M_{GC}$=0.005, 0.01 and 0.02 respectively. The initial concentrations and half-mass radii
of these models were varied in the same way as for the no-IMBH models described above. All clusters in this paper were isolated, however we plan to add external
tidal fields in subsequent papers when we compare the internal mass function of stars at different radii with observations.

Since our simulations contain fewer stars than the actual globular clusters and have different half-mass radii at the end of the simulations, we need to scale our
simulations to match the masses and sizes of observed globular clusters. Our scaling procedure is the same as that used by    
\citet{baumgardtetal2003a} who fitted the massive globular cluster G1 in M31 by a set of $N$-body simulations, and \citet{jalalietal2012} who
fitted $\omega$ Cen by a set of $N$-body models. The basic assumption of the scaling is that since the simulated star clusters are isolated, they evolve only due to
stellar evolution and two-body relaxation. Hence the simulations can be scaled to star clusters of different mass or radius as long as the scaling is done
in such a way that the overall relaxation time remains constant.
Using the definition of the half-mass relaxation time given by \citet{spitzer1987}, this implies
\begin{equation}
\frac{r_{NB}}{r_{GC}} = \left(\frac{M_{GC}}{M_{NB}}\right)^{1/3} \left(\frac{ln \gamma N_{NB}}{ln \gamma N_{GC}} \right)^{2/3}
\label{radscale}
\end{equation}
where $M$ is the mass of a cluster, $r$ its half-mass radius, $N=M/\!<\!m\!>$ the number of cluster stars, and the subscripts NB and GC refer
respectively to a star cluster from our grid of $N$-body simulations and an observed globular cluster that we want to model.
$\gamma$ is a constant in the Coulomb logarithm which we assume to be equal to 0.11 \citep{gierszheggie1994}.

We determine the projected half-light radius of a globular cluster by integrating the observed surface density profile up to the outermost radius
for which data is available. For each simulated cluster, we then determine iteratively the scaling factor $f_r=r_{NB}/r_{GC}$ that is necessary so that
the cluster from the $N$-body simulation has the same projected half-light radius inside the same limiting radius as the observed globular cluster if put
at the same distance as the observed globular cluster. For the calculation of the surface brightness
profiles of the simulated clusters, we converted the bolometric luminosities of NBODY6 to $V$-band luminosities using the conversion formulae given by \citet{eggletonetal1989}.
After determining the radial scaling factor $f_r$, we determine the corresponding mass scaling factor from eq. \ref{radscale} and then multiply the velocities of the stars
in the $N$-body simulation by a factor
\begin{equation}
 f_v = \left( \frac{r_{NB}}{r_{GC}} \right)^{1/2} \left( \frac{M_{GC}}{M_{NB}} \right)^{1/2} \;\; ,
\label{velscale}
\end{equation}
where the first term on the right-hand side is due to the change in radius and the second term is due to the change in cluster mass.
After scaling the velocities, we calculate the surface density, and line-of-sight and proper motion velocity dispersion profiles for the simulated clusters.
In order to improve the statistical significance of our results in the cluster centers, we
use the infinite projection method of \citet{mashchenkosills2005} when calculating surface density and velocity dispersion profiles. For the
velocity dispersion profiles we mimic the magnitude limits
of the observations by using only stars brighter than the main-sequence turn-off to determine the line-of-sight velocity dispersion profile. To compare with the
proper motion data of \citet{watkinsetal2015a} we use all stars brighter than 1 mag below the turnoff magnitude.
The resulting velocity dispersion profiles differ due to mass segregation, however the differences are typically less than 5\%.
\begin{figure}
\begin{center}
\includegraphics[width=8cm]{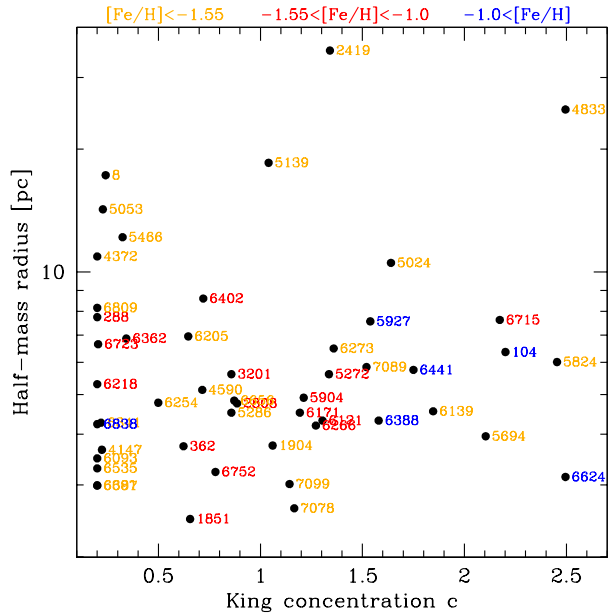}
\end{center}
\caption{Location of the best-fitting no-IMBH model for each globular cluster within the simulated grid of models. Individual globular clusters are marked
by their NGC numbers. Small half-mass radii imply small initial
relaxation times and therefore dynamically more advanced globular clusters. Most globular clusters can be fitted by models starting from initial
half-mass radii between 3 to 7 pc corresponding to initial relaxation times between 0.5 to 2 Gyr.}
\label{fitgrid}
\end{figure}

In order to increase the number of models that can be compared with each globular cluster, we assume that the properties of the final cluster
change linearly with the initial concentration $c$, the logarithm of the initial half-mass radius $\log r_h$ and the IMBH mass fraction $M_{BH}/M_{GC}$
and interpolate between our grid points. In total we use 300 interpolation values for each grid dimension and determine the best-fitting model 
to the observed surface brightness and velocity dispersion profile by means of a $\chi^2$ test.

Fig.~\ref{fitgrid} shows the location of the best-fitting no-IMBH model for each globular cluster within the simulated grid of models. In this figure a small initial half-mass radius
$r_h$ implies a small initial relaxation time and therefore a more dynamically advanced globular cluster. It does not necessarily imply that a  
cluster actually started with a small half-mass radius, although relaxation time and half-mass radius are correlated with each other. Most 
clusters can be fitted with clusters starting with
half-mass radii around 5~pc, implying initial relaxation times of $T_{RH} \approx 1$ Gyr. The best-fitting
models of most globular clusters are located within our grid boundaries, however for nine globular clusters we need models with the lowest
modeled King concentration parameter of $c=0.2$ to fit their surface density profiles. A look at Figs.~\ref{fig1a} to \ref{fig13a} shows that
we nevertheless usually obtain very good fits to their surface density and velocity dispersion profiles, so the low initial concentrations are not of
immediate concern. They might however be an indication that either the surface density profiles of these clusters are influenced by
the tidal field of the Milky Way or ongoing mass loss, processes which are not included in our simulations. Indeed most of these clusters have
small galactocentric radii ($R_G<5$ kpc) where tidal effects should be most important. Alternatively, a compact cluster of stellar mass
black holes might prevent the cores of these clusters from collapsing \citep{morscheretal2013, lkb13}. Indeed, this possibility has been suggested by
\citet{mackeyetal2007} to explain the large core radii of young star clusters in the LMC and more recently by \citet{peutenetal2016} to explain the
absence of mass segregation in NGC 6101. Additional simulations will be necessary to distinguish between these possibilities.

\subsection{Validation}

\begin{figure}
\begin{center}
\includegraphics[width=8cm]{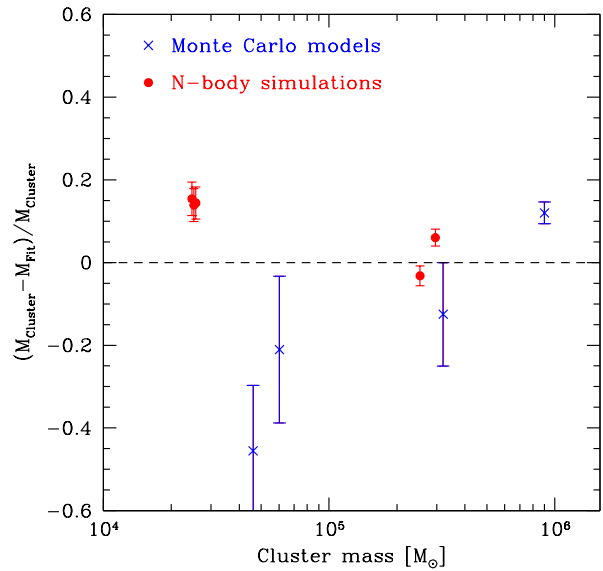}
\end{center}
\caption{Comparison of the masses derived from our grid of $N$-body simulations with the true masses of star clusters in $N$-body simulations (red circles) and the masses of Galactic globular
clusters derived by fitting results of Monte-Carlo simulations (blue crosses). Our mass estimates reproduce the true masses of star clusters in $N$-body simulations to within
10\% and are within 20\% of the masses of Galactic globular clusters derived from Monte-Carlo simulations.}
\label{vmasses}
\end{figure}

In order to test how well our fitting method can reproduce star cluster masses from their surface density and velocity dispersion profiles we
apply our models to the $N$-body simulations uf13 and uf14 from \citet{lamersetal2013} and models D1 and D2 from the {\tt DRAGON} simulation \citep{wangetal2016}.
We use four snapshots of simulations uf13 and uf14 between $T=11.5$ and $T=12.5$ Gyr and one snapshot of models D1 and D2 at $T=12$ Gyr and calculate 
the surface density and velocity dispersion profile using all stars that are still bound to the clusters
at these times. We then apply our fitting method to the four clusters. Fig.~\ref{vmasses} compares
the derived masses with the true masses of the simulated clusters (red circles). By the time the snapshots are created, the simulated clusters have
lost between 12\% to 75\% of their initial mass and for some of the clusters the mass function has already evolved significantly away 
from a Kroupa mass function. Nevertheless our fitting method reproduces the cluster masses to within 10\%. It performs slightly better for the dynamically
less evolved clusters of the {\tt DRAGON} simulation and less well for the highly evolved clusters from \citet{lamersetal2013}.

As a second check, we compare results of our fitting method with the results of Monte-Carlo simulations aimed to reproduce the
luminosity and velocity dispersion profiles and the luminosity function of stars in a number of globular clusters. The Monte Carlo results were 
published by \citet{gierszheggie2011} (for NGC 104), \citet{heggiegiersz2008} (NGC 6121), \citet{gierszheggie2009} (NGC6397) and \citet{heggiegiersz2014} (NGC 6656).
For all clusters we adopt the same distances as assumed by Giersz \& Heggie and use only the velocity dispersion data which Giersz \& Heggie used for each cluster.
Fig. \ref{vmasses} compares the masses which we derive from $N$-body models with those found in the Monte-Carlo simulations. We can reproduce the masses from the 
Monte-Carlo simulations to within $\sim$20\%. The deviations are again larger for the dynamically more evolved clusters NGC 6121 and NGC 6397 and better for the more 
massive clusters NGC 104 and NGC 6656. The reason for the larger deviation of the Monte Carlo models compared to the $N$-body simulations is probably the small number of radial
velocity data points of the observed clusters, which leave large freedom in the mass profiles and total cluster masses.
We conclude that our models can reproduce cluster masses to within 10\% for clusters that have a well determined radial velocity dispersion profile.
Better mass estimates will probably require knowledge of the internal mass function of the cluster stars in addition to the clusters' velocity dispersion and surface density
profile.

\begin{table}
\caption{Input parameters for the studied globular clusters. The sources for the distances are: F99: \citet{ferraroetal1999}, V07: \citet{valentietal2007}, D11: \citet{dicri2011}, Z98: \citet{zinnbarnes1998}, H96: \citet[2010 edition]{harris1996}, tw: This work}
\begin{tabular}{ll@{}rc@{$\;$}c@{$\,$}r@{$\,\,$}c}
\hline
 \multicolumn{1}{c}{Name} & \multicolumn{1}{c}{Alt.}& \multicolumn{1}{c}{$V$} & \multicolumn{1}{c}{$\Delta V$} & \multicolumn{1}{c}{Age} & \multicolumn{1}{c}{Dist.} & Dist. \\
   & \multicolumn{1}{c}{name} & \multicolumn{1}{c}{[mag]} & \multicolumn{1}{c}{[mag]} & \multicolumn{1}{c}{[Gyr]} & \multicolumn{1}{c}{[kpc]} & Source \\
\hline
NGC 104 &  47 Tuc       &  4.07 & 0.11 & 11.75 &  3.95 & tw \\
NGC 288 &               &  8.16 & 0.07 & 11.50 &  8.80 & tw \\
NGC 362 &               &  6.55 & 0.16 & 10.75 &  8.85 & tw \\
NGC 1851 &               &  7.24 & 0.09 & 11.00 & 10.40 & tw \\
NGC 1904 &  M 79         &  7.99 & 0.19 & 11.70 & 13.27 & F99\\
NGC 2419 &               & 10.48 & 0.15 & 12.75 & 87.50 & D11\\
NGC 2808 &               &  6.33 & 0.09 & 11.00 &  9.50 & tw \\
NGC 3201 &               &  6.88 & 0.20 & 11.50 &  4.90 & F99\\
NGC 4147 &               & 10.38 & 0.11 & 12.25 & 18.20 & F99\\
NGC 4372 &               &  7.23 & 0.01 & 12.00 &  6.30 & F99\\
NGC 4590 &  M 68         &  8.15 & 0.22 & 12.00 & 10.59 & F99\\
NGC 4833 &               &  6.91 & 0.20 & 12.50 &  6.76 & F99\\
NGC 5024 &  M 53         &  7.71 & 0.10 & 12.25 & 17.90 & H96\\
NGC 5053 &               &  7.71 & 0.10 & 12.25 & 17.20 & F99\\
NGC 5139 &  $\omega$ Cen &  3.53 & 0.11 & 12.00 &  5.00 & tw \\
NGC 5272 &  M 3          &  6.40 & 0.16 & 11.75 & 10.06 & F99\\
NGC 5286 &               &  7.20 & 0.12 & 12.50 & 11.70 & H96\\
NGC 5466 &               &  9.46 & 0.30 & 12.50 & 16.90 & F99\\
NGC 5694 &               & 10.02 & 0.14 & 12.75 & 37.33 & F99\\
NGC 5824 &               &  8.83 & 0.19 & 13.00 & 31.80 & F99\\
NGC 5904 &  M 5          &  5.83 & 0.16 & 11.50 &  6.40 & tw \\
NGC 5927 &               &  7.74 & 0.39 & 10.75 &  8.00 & tw \\
NGC 6093 &  M 80         &  7.35 & 0.13 & 11.40 &  9.73 & F99\\
NGC 6121 &  M 4          &  5.63 & 0.09 & 11.50 &  2.14 & F99\\
NGC 6139 &               &  8.95 & 0.13 & 12.00 & 10.40 & Z98\\
NGC 6171 &  M 107        &  8.18 & 0.31 & 12.00 &  6.09 & F99\\
NGC 6205 &  M 13         &  5.80 & 0.10 & 12.00 &  7.60 & F99\\
NGC 6218 &  M 12         &  6.92 & 0.28 & 13.00 &  5.22 & F99\\
NGC 6254 &  M 10         &  6.42 & 0.38 & 11.75 &  4.71 & F99\\
NGC 6266 &  M 62         &  6.45 & 0.12 & 11.40 &  6.55 & tw \\
NGC 6273 &  M 19         &  6.80 & 0.05 & 12.75 &  8.24 & V07\\
NGC 6341 &  M 92         &  6.51 & 0.06 & 12.75 &  8.10 & tw \\
NGC 6362 &               &  7.67 & 0.10 & 12.50 &  7.60 & H96\\
NGC 6388 &               &  6.76 & 0.13 & 11.75 & 11.00 & tw \\
NGC 6397 &               &  5.77 & 0.18 & 13.00 &  2.40 & tw \\
NGC 6402 &  M 14         &  7.66 & 0.08 & 11.50 &  9.30 & H96\\
NGC 6441 &               &  7.16 & 0.11 & 11.00 & 13.49 & V07\\
NGC 6535 &               & 11.14 & 0.57 & 12.75 &  7.28 & H96\\
NGC 6624 &               &  7.78 & 0.13 & 11.25 &  8.43 & V07\\
NGC 6656 &  M 22         &  5.07 & 0.07 & 12.50 &  2.66 & tw \\
NGC 6681 &  M 70         &  7.98 & 0.15 & 12.75 &  9.89 & F99\\
NGC 6715 &  M 54         &  7.47 & 0.10 & 11.75 & 23.50 & tw \\
NGC 6723 &               &  7.11 & 0.17 & 12.50 &  8.20 & V07\\
NGC 6752 &               &  5.52 & 0.17 & 12.50 &  3.90 & tw \\
NGC 6809 &  M 55         &  6.63 & 0.24 & 13.00 &  5.75 & F99\\
NGC 6838 &  M 71         &  7.84 & 0.49 & 11.00 &  3.86 & F99\\
NGC 7078 &  M 15         &  6.13 & 0.10 & 12.75 &  9.90 & tw \\
NGC 7089 &  M 2          &  6.43 & 0.03 & 11.75 & 11.50 & H96\\
NGC 7099 &  M 30         &  7.25 & 0.24 & 13.00 &  8.67 & F99\\
Terzan 8 &               & 12.11 & 0.32 & 13.00 & 26.73 & F99\\
\hline
\end{tabular}
\label{gcinput}
\end{table}

\section{Globular cluster data}

We first determined the radial velocity dispersion profiles of Galactic globular clusters from individual radial velocity measurements of their member stars published
in the literature. To this end, we searched the astronomical literature for published radial velocity
measurements, excluding small data sets with less than $\approx 20$ stars. In total we found 95 publications containing about 25500 individual 
radial velocities of stars in 45 clusters. 
About one third of the radial velocity measurements were from the three, recent large-scale surveys of \citet{laneetal2011}, \citet{lardoetal2015} and 
\citet{kimmigetal2015}, which each contain radial velocity information for several thousand stars. The rest of the data comes from smaller data sets.
For nine globular clusters we also included radial velocities from the APOGEE survey \citep{majewskietal2015}, which has measured abundances and 
radial velocities for over 150,000 red giants, including several hundred stars in globular clusters.
Information on the papers used as input for calculating the radial velocity dispersion profiles can be found in 
Table~\ref{sourcestab}\footnote{The radial velocity dispersion profiles can be downloaded from \href{https://people.smp.uq.edu.au/HolgerBaumgardt/globular/}{https://people.smp.uq.edu.au/HolgerBaumgardt/globular/}}.

For each individual set of radial velocities, we first calculated the average cluster velocity
using the method of \citet{pryormeylan1993} and using all stars which roughly fall within the radial velocity range of the cluster.
We then subtracted the average cluster velocity from the individual measurements and merged all radial velocity data sets into a master catalogue
for each cluster, containing the positions, radial velocities and radial velocity errors of all stars. We then use the stellar positions to
identify stars with multiple measurements and calculate a weighted mean radial velocity and corresponding error for each star with multiple measurements. 
Stars for which the individual radial velocity measurements show a too strong deviation from the mean were rejected as binaries.
After removing binary stars, we put the stars into radial bins and calculated the radial velocity dispersion $\sigma_{bin}$ by 
determining the maximum of the likelihood function 
\begin{equation}
 \log L = -\frac{1}{2} \sum_{i=1}^N \ln \left(\sigma_{bin}^2+e_i^2\right) + \sum_{i=1}^N\frac{v_i^2}{\sigma_{bin}^2+e_i^2}  \;\; . 
\end{equation} 
based on all stars in a bin.
Here $v_i$ and $e_i$ are the radial velocity and its respective error of each individual star. The 1$\sigma$ lower and upper uncertainties of the 
velocity dispersion were calculated by determining the velocity dispersion where the likelihood is less than 0.5 the maximum value in each direction.
After the velocity dispersion of a radial bin was determined, we calculated the deviation of each star from the cluster mean according to 
\begin{equation}
 \chi^2 = \frac{v^2_i}{\sigma^2+e_i^2}
\end{equation}
and rejected all stars as binaries or background stars that deviated more than three standard deviations from the mean. We repeated the above
procedure for each bin until we found a stable value for the velocity dispersion and the list of member stars. Depending on the number of radial 
velocity measurements available
for a cluster, we used between 20 to 250 stars per bin to calculate the radial velocity dispersion. In order to calculate the radial velocity 
dispersion profile, we used the positions determined by \citet{goldsburyetal2013} as cluster centers, except for NGC 1904, NGC 5694, NGC 5824 
and NGC 6266 where we used the centers determined by \citet{lutzgendorfetal2013}. This was necessary in order to get a radial velocity 
dispersion profile centered on the same position as the IFU data published by \citet{lutzgendorfetal2013}. For clusters that contain a significant number of stars at large distances
from the cluster center, we used proper motions from the PPMXL catalogue \citep{roeseretal2010} to help separate cluster members from non-members.
PPMXL data was only used to separate members from non-members for stars more than a few hundred arcsec away from the cluster center since
for stars closer to the center PPMXL proper motions were either not available or were found to be unreliable, presumably due to the strong
crowding of stars towards the cluster centers.

In addition to radial velocity dispersion data, we also used velocity dispersion data based on 
individual stellar proper motions to constrain the cluster kinematics. Most of the proper motion dispersion profiles were taken from 
\citet{watkinsetal2015a}, who published velocity dispersion profiles for 21 clusters. We excluded NGC 7099 since the
proper motion dispersion profile from \citet{watkinsetal2015a} disagrees significantly from the radial velocity dispersion profile calculated
in this paper for any reasonable cluster distance. As discussed by \citet{watkinsetal2015a} this might be due to the small number of stars
which have measured proper motions in this cluster. We finally used velocity dispersion measurements based on integral-field unit (IFU) spectroscopy. IFU spectroscopy
was available for 9 clusters (NGC 1851, 1904, 2808, 5286, 5694, 5824, 6093, 6266 and NGC 6388).

The surface brightness profiles were taken mainly from \citet{trageretal1995}. If available for a cluster, we replaced the Trager et al. profile
in the cluster center with HST surface brightness profiles published by \citet{noyolagebhardt2006}. For a few clusters we took the surface density 
profiles from other literature sources. These cases are listed in Tab.~\ref{sourcestab} in the Appendix. For NGC 5927 the surface density profile
calculated by \citet{trageretal1995} has a bump in the center that is impossible to reproduce by our modeling. Since no other cluster shows
such a feature, the surface density profile might be influenced by a few bright stars in the center. We therefore calculated a surface brightness
profile based on the number counts of bright stars published by the ACS Survey of Galactic Globular Clusters \citep{sarajedinietal2007}.
The same was done for NGC 4833 where we combined data published by \citet{melbourneetal2000} with the ACS data to calculate the surface density profile.

16 clusters from our list have accurate proper motion dispersion profiles
and also accurate enough radial velocity dispersion profiles so that their distances can be determined by $\chi^2$ minimization
of a simultaneous fit of our models to both profiles. From the fits we are able to measure their distances to an accuracy of
between 50 to 450 pc and the distances are given in Table \ref{tabresult}. For the remaining clusters, the 
distances were taken mainly from \citet{ferraroetal1999}, who determined globular cluster distance moduli by CMD fitting.
For clusters not studied by \citet{ferraroetal1999}, we took the distances from recent literature values. The adopted distances
are listed in Table~\ref{gcinput} together with the cluster ages and the calculated $V$-band magnitudes. The apparent $V$-band magnitudes and
errors are calculated by taking the average of the apparent magnitudes given in \citet{harris1996}, \citet{mclaughlinvandermarel2005}, 
\citet{dalessandroetal2012} and the integrated magnitudes determined in this work from the fit of our models to the surface brightness profiles. Cluster ages were taken 
from \citet{vandenbergetal2013}, or, if not available, from literature data. We finally took the cluster metallicities from
the recent compilation by \citet{carrettaetal2009b} and the cluster reddenings from \citet{harris1996}.

\begin{table}
\caption{Derived parameters of the studied globular clusters}
\begin{tabular}{@{}l@{\hspace*{0.3cm}}c@{\hspace*{0.3cm}}c@{\hspace*{0.2cm}}r@{\hspace*{0.3cm}}r@{$\,\pm\,$}l@{}}
\hline
Name & $\chi^2_{red}$ & Mass & \multicolumn{1}{c}{M/L ratio} & \multicolumn{2}{c}{Distance}\\
 & & \multicolumn{1}{c}{[M$_\odot$]} & & \multicolumn{2}{c}{[pc]} \\
\hline
NGC 104 & $2.01$ & $7.00 \pm 0.06 \cdot 10^5$ & $1.99 \pm 0.20$ & 3950 &  50 \\
NGC 288 & $1.43$ & $8.76 \pm 0.26 \cdot 10^4$ & $2.23 \pm 0.15$ & 8800 & 400 \\
NGC 362 & $0.76$ & $3.21 \pm 0.06 \cdot 10^5$ & $1.73 \pm 0.26$ & 8850 & 300 \\
NGC 1851 & $1.81$ & $2.99 \pm 0.05 \cdot 10^5$ & $2.40 \pm 0.20$ & 10400 & 200 \\
NGC 1904 & $1.95$ & $2.20 \pm 0.18 \cdot 10^5$ & $2.23 \pm 0.43$ \\
NGC 2419 & $2.60$ & $8.15 \pm 1.19 \cdot 10^5$ & $1.54 \pm 0.22$ \\
NGC 2808 & $2.13$ & $8.29 \pm 0.06 \cdot 10^5$ & $1.96 \pm 0.16$ & 9500 & 150 \\
NGC 3201 & $1.51$ & $1.58 \pm 0.11 \cdot 10^5$ & $2.20 \pm 0.43$ \\
NGC 4147 & $1.60$ & $5.32 \pm 1.71 \cdot 10^4$ & $2.45 \pm 0.32$ \\
NGC 4372 & $0.32$ & $2.20 \pm 0.25 \cdot 10^5$ & $1.67 \pm 0.19$ \\
NGC 4590 & $0.95$ & $8.45 \pm 1.71 \cdot 10^4$ & $1.39 \pm 0.65$ \\
NGC 4833 & $0.74$ & $2.66 \pm 0.39 \cdot 10^5$ & $1.59 \pm 0.33$ \\
NGC 5024 & $0.61$ & $3.83 \pm 0.51 \cdot 10^5$ & $1.60 \pm 0.84$ \\
NGC 5053 & $0.54$ & $5.37 \pm 1.32 \cdot 10^4$ & $1.58 \pm 0.54$ \\
NGC 5139 & $2.56$ & $2.95 \pm 0.02 \cdot 10^6$ & $2.54 \pm 0.26$ & 5000 &  50 \\
NGC 5272 & $1.85$ & $5.00 \pm 0.43 \cdot 10^5$ & $1.98 \pm 0.37$ \\
NGC 5286 & $1.09$ & $4.61 \pm 0.23 \cdot 10^5$ & $1.51 \pm 0.18$ \\
NGC 5466 & $0.92$ & $6.43 \pm 1.47 \cdot 10^4$ & $1.60 \pm 0.56$ \\
NGC 5694 & $0.97$ & $4.22 \pm 0.45 \cdot 10^5$ & $2.79 \pm 0.42$ \\
NGC 5824 & $0.32$ & $8.28 \pm 0.55 \cdot 10^5$ & $2.25 \pm 0.42$ \\
NGC 5904 & $1.08$ & $3.08 \pm 0.04 \cdot 10^5$ & $1.74 \pm 0.26$ & 6400 & 200 \\
NGC 5927 & $1.96$ & $3.45 \pm 0.03 \cdot 10^5$ & $2.19 \pm 0.42$ & 8000 & 400 \\
NGC 6093 & $1.46$ & $3.37 \pm 0.16 \cdot 10^5$ & $2.18 \pm 0.28$ \\
NGC 6121 & $0.89$ & $1.01 \pm 0.03 \cdot 10^5$ & $1.70 \pm 0.15$ \\
NGC 6139 & $0.55$ & $5.31 \pm 1.22 \cdot 10^5$ & $2.59 \pm 0.61$ \\
NGC 6171 & $0.96$ & $9.62 \pm 1.04 \cdot 10^4$ & $2.22 \pm 0.69$ \\
NGC 6205 & $2.03$ & $5.00 \pm 0.42 \cdot 10^5$ & $2.06 \pm 0.33$ \\
NGC 6218 & $0.70$ & $1.03 \pm 0.12 \cdot 10^5$ & $1.51 \pm 0.40$ \\
NGC 6254 & $1.05$ & $2.26 \pm 0.29 \cdot 10^5$ & $1.99 \pm 0.72$ \\
NGC 6266 & $1.58$ & $9.31 \pm 0.09 \cdot 10^5$ & $2.54 \pm 0.28$ & 6550 & 140 \\
NGC 6273 & $0.08$ & $9.21 \pm 1.62 \cdot 10^5$ & $2.83 \pm 0.45$ \\
NGC 6341 & $0.74$ & $3.05 \pm 0.04 \cdot 10^5$ & $2.06 \pm 0.12$ & 8100 & 150 \\
NGC 6362 & $1.26$ & $1.44 \pm 0.05 \cdot 10^5$ & $2.64 \pm 0.26$ \\
NGC 6388 & $1.11$ & $1.24 \pm 0.01 \cdot 10^6$ & $2.11 \pm 0.26$ & 11000 & 450 \\
NGC 6397 & $1.09$ & $9.40 \pm 0.32 \cdot 10^4$ & $2.33 \pm 0.39$ & 2400 &  60 \\
NGC 6402 & $1.43$ & $7.63 \pm 1.19 \cdot 10^5$ & $2.17 \pm 0.37$ \\
NGC 6441 & $1.55$ & $1.86 \pm 0.02 \cdot 10^6$ & $2.30 \pm 0.24$ \\
NGC 6535 & $2.28$ & $5.96 \pm 0.59 \cdot 10^4$ & $14.29 \pm 7.93$ \\
NGC 6624 & $1.70$ & $2.42 \pm 0.07 \cdot 10^5$ & $2.33 \pm 0.29$ \\
NGC 6656 & $0.93$ & $3.21 \pm 0.04 \cdot 10^5$ & $2.15 \pm 0.14$ & 2660 & 100 \\
NGC 6681 & $1.31$ & $1.72 \pm 0.04 \cdot 10^5$ & $2.62 \pm 0.36$ \\
NGC 6715 & $5.07$ & $1.62 \pm 0.03 \cdot 10^6$ & $2.18 \pm 0.20$ & 23500 & 300 \\
NGC 6723 & $0.26$ & $1.96 \pm 0.40 \cdot 10^5$ & $2.06 \pm 0.41$ \\
NGC 6752 & $0.71$ & $2.34 \pm 0.04 \cdot 10^5$ & $2.60 \pm 0.41$ & 3900 & 100 \\
NGC 6809 & $3.34$ & $1.78 \pm 0.15 \cdot 10^5$ & $2.25 \pm 0.52$ \\
NGC 6838 & $1.43$ & $4.60 \pm 0.61 \cdot 10^4$ & $2.43 \pm 1.18$ \\
NGC 7078 & $1.72$ & $5.01 \pm 0.06 \cdot 10^5$ & $1.27 \pm 0.12$ & 9900 & 200 \\
NGC 7089 & $0.46$ & $7.64 \pm 0.51 \cdot 10^5$ & $2.13 \pm 0.15$ \\
NGC 7099 & $0.58$ & $1.21 \pm 0.10 \cdot 10^5$ & $1.37 \pm 0.32$ \\
Terzan 8 & $0.49$ & $5.37 \pm 2.34 \cdot 10^4$ & $4.36 \pm 2.96$ \\
\hline
\end{tabular}
\label{tabresult}
\end{table}

\section{Results}

Figs. \ref{fig1a} to \ref{fig13a} compare our best-fitting profiles with the observed velocity dispersion and surface density profiles of globular clusters.
Except for $\omega$ Cen and NGC 6715 all profiles shown are the no-IMBH models.
As can be seen we usually obtain very good fits to the observed profiles. The surface brightness profiles of our best-fitting clusters are generally
within 20\% of the observed surface brightness, despite the fact that the observed surface brightness profiles vary by up to 6 orders of magnitude in some clusters. Only
beyond several hundred arcsec, some clusters show larger differences in their surface density profiles. This could be due to the influence of the Galactic tidal field which was not
taken into account in our simulations, but might also be a result of observational uncertainties since a few hundred arcsec from the cluster center the surface 
density of many globular clusters is already significantly below the background density of stars, making the determination of the outer surface density profiles uncertain.  
The differences with the measured velocity dispersion profiles are also usually less than 1~km/sec and for most clusters within the 
observational uncertainties. The only clusters which cannot be well modeled by the no-IMBH models
are $\omega$ Cen and M54 (NGC 6715). For these clusters the observed velocity dispersion profile is significantly above our predictions in the center and
below in the outer parts. This could be due to an unseen mass concentration in the center and we will discuss these clusters in greater detail in sec.~\ref{sec:imbh}
when we investigate the possible presence of IMBHs in globular clusters.
\begin{figure}
\begin{center}
\includegraphics[width=8.0cm]{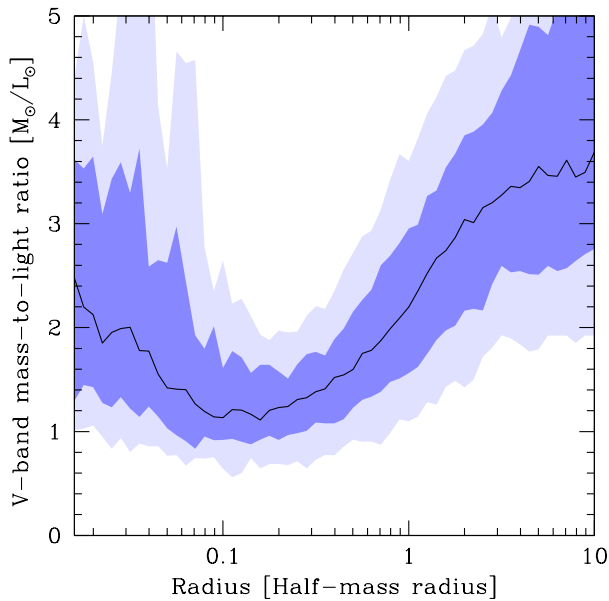}
\end{center}
\caption{Local V band mass-to-light ratios as a function of the distance to the cluster center expressed in units of the half-mass radius.  The average $M/L$ ratio of
all 50 clusters is shown by a solid line. Dark blue and light blue shaded regions mark the values of the $M/L$ ratio which contain 68\% and 95\% of all clusters. The $M/L$
ratio profiles of the clusters follow a U-shaped curve due to mass segregation, which concentrates high-mass compact remnants and giant stars in the cluster center 
and pushes low-mass main sequence stars towards the cluster outskirts.} 
\label{mlratiovar}
\end{figure}

Fig. \ref{mlratiovar} depicts the V-band mass-to-light ratio profiles which we derive from our fits. The average mass-to-light ratio of all 50 clusters as a function of distance to the
cluster center is shown by a solid line and the regions in $M/L$ ratio that contain 68\% and 95\% of all clusters are shown by dark and light blue areas respectively. In order 
to better compare individual clusters, we have divided the distances to the cluster centers by the half-mass radius of each cluster. It can be seen that the $M/L$ ratios
have a minimum between 0.1 to 0.2 half-mass radii. This minimum is due to the mass segregation of giant stars and high-mass main sequence stars towards the cluster center. Since giant
stars dominate the cluster light but contain only a small fraction of the cluster mass, the $M/L$ ratio decreases in the center. Inside
of 0.1 half-mass radii the $M/L$ ratios rise again since compact remnants like high-mass white dwarfs, neutron stars and black holes have masses even higher than the giant stars
and are therefore more strongly concentrated towards the cluster center. The $M/L$ ratios also increase towards the outer cluster parts since low-mass main sequence stars
are pushed out of the cluster due to mass segregation. We also find that the importance of mass segregation depends on the relaxation time of a cluster. In clusters with 
very large relaxation times like NGC 2419, the $M/L$ ratio changes by less than 30\% between the center and the cluster halo. In contrast, for strongly mass segregated clusters the variation of the
$M/L$ ratio can reach a factor of 4 between the core region and the cluster outskirts. This agrees with recent results of Monte Carlo simulations by \citet{bianchinietal2016}, 
who found that the amount of mass segregation tightly correlates with the dynamical state of the cluster.

Table \ref{tabresult} presents a summary of our results. It gives the name of the cluster, the reduced $\chi^2$ value from fitting the velocity dispersion and surface brightness profiles, 
the derived
cluster mass and its error, the global $M/L$ ratio and its error, and the best-fitting cluster distance and its error for those clusters where we derived
cluster distances ourselves. Errors in the $M/L$ ratio were calculated from the errors in cluster mass and cluster luminosity but do not
include uncertainties in the cluster distances. The average V-band $M/L_V$ ratio for our whole cluster sample is $M/L_V=1.98 \pm 0.03$ $M_\odot/L_\odot$ and
$M/L_V=1.98 \pm 0.04$ $M_\odot/L_\odot$ if we restrict ourselves to clusters that have more than 200 radial velocity measurements and mass-to-light ratios
with relative errors less than 30\%. If we split the more accurate cluster sample into two sub-samples depending on cluster metallicity, we derive a mean 
V-band mass-to-light ratio of $M/L_V=1.88 \pm 0.06$
$M_\odot/L_\odot$ for the metal-poor clusters with [Fe/H]$<-1.5$ and $M/L_V=2.07 \pm 0.06$ $M_\odot/L_\odot$ for the metal-rich clusters. This increase of the average
mass-to-light ratio with metallicity is in general agreement with predictions from stellar evolution theory.

\begin{figure}
\begin{center}
\includegraphics[width=8.0cm]{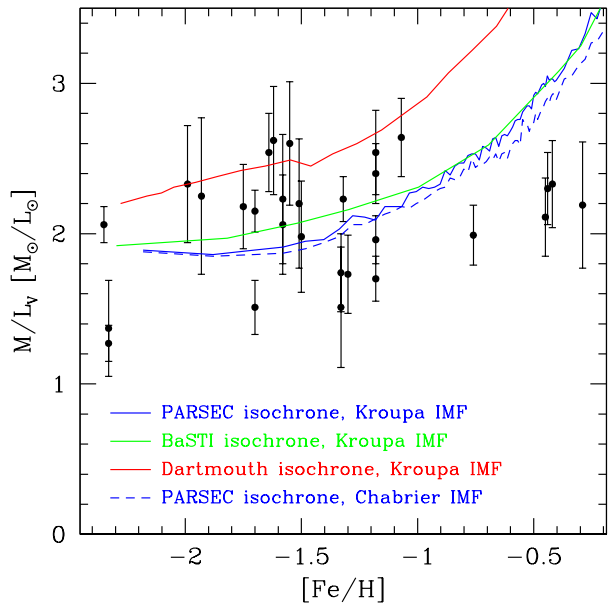}
\end{center}
\caption{V band mass-to-light ratios derived in this work as a function of metallicity. Solid lines show the predicted mass-to-light ratios for a Kroupa IMF according to the
PARSEC (blue), BaSTI (green), and Dartmouth (red) isochrones for an age of $T=12.5$ Gyr. The dashed blue line shows the predicted $M/L$ ratio from the PARSEC isochrones
for a Chabrier IMF. Except for the Dartmouth isochrones, our derived $M/L$ ratios agree well with the theoretical predictions for either a Kroupa or Chabrier IMF.}
\label{mlratio}
\end{figure}

Fig. \ref{mlratio} compares the global $M/L$ ratios derived here with the predictions of stellar evolution models. Shown are predicted $M/L$ ratios
from PARSEC \citep{bressanetal2012}, $\alpha$ enhanced BaSTI \citep{pietrinfernietal2006} and $\alpha$ enhanced Dartmouth isochrones \citep{dotteretal2008}. 
The theoretical $M/L_V$ values were calculated assuming 
a \citet{kroupa2001} IMF with mass limits of 0.1 and 100 M$_\odot$, the \citet{kaliraietal2008} initial-final mass ratio for white dwarfs and a 
10\% retention fraction of neutron stars and black holes in the clusters. 
For the PARSEC isochrones, we also calculated $M/L_V$ ratios for stars distributed according to a \citet{chabrier2003} IMF between mass limits of 0.1 and 100 $M_\odot$.
Since the BaSTI isochrones only give luminosities for stars with masses larger than 0.5 M$_\odot$, we used PARSEC luminosities for less massive stars.
For clarity, we show only clusters with more than 200 radial velocity measurements and mass-to-light ratios
with relative errors less than 30\% in Fig.\ \ref{mlratio}, however the full cluster sample has a very similar distribution. It can be seen that the derived mass-to-light ratios are in 
general agreement with the PARSEC and BaSTI isochrones, especially at low metallicity. The agreement is less good for the Dartmouth isochrones, however these isochrones have
a less  detailed treatment of giant star evolution than either the BaSTI or PARSEC isochrones. Since giant stars dominate the cluster light we regard the predictions of
the BaSTI or PARSEC isochrones as more reliable. 

\citet{straderetal2011} found a decrease of the $M/L$ ratio down to about $M/L_V \approx 1$ $M_\odot/L_\odot$ for solar metallicity for globular clusters in M31, which they attributed
to a systematic change of the IMF with metallicity. We do not see 
a decrease of the $M/L$ ratio with increasing metallicity. A possible reason could be that \citet{straderetal2011} fitted single-mass King models
to derive the global velocity dispersion from the measured central one. This will produce a bias in the derived masses if clusters
are mass segregated. 

\begin{figure}
\begin{center}
\includegraphics[width=8.0cm]{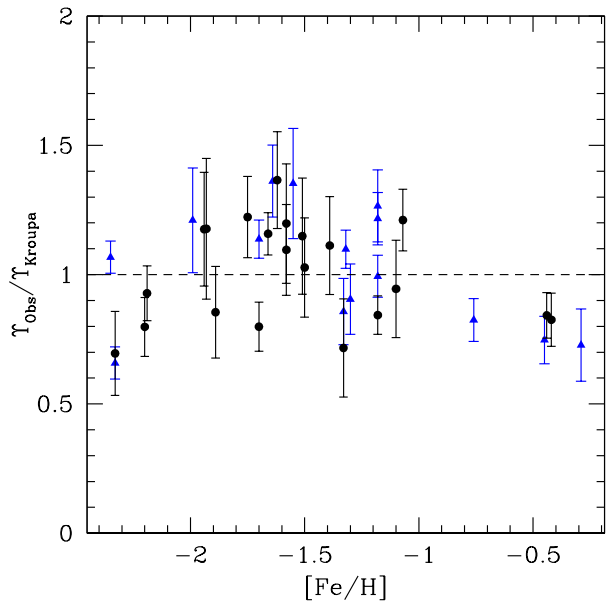}
\end{center}
\caption{Ratio of the measured $M/L$ ratios to the $M/L$ ratios predicted by the PARSEC isochrones for clusters with a Kroupa IMF at the measured ages of the clusters as a function
of cluster metallicity. Blue triangles mark clusters for which the distances were determined in this work. The average $\Upsilon_{Obs}/\Upsilon_{Kroupa}$ ratio is close to unity,
indicating that most clusters have mass functions compatible with a Kroupa IMF.}
\label{mlratio2}
\end{figure}

In order to better compare the derived $M/L$ ratios with predictions of stellar evolution models, we depict in Fig. \ref{mlratio2} the ratio of the observed $M/L$ ratio 
$\Upsilon_{Obs} = M/L_V$  to $\Upsilon_{Kroupa}$, the $M/L_V$ ratio predicted by the PARSEC isochrones for clusters with a Kroupa IMF at the measured age of each individual cluster.
The average $\Upsilon_{Obs}/\Upsilon_{Kroupa}$ ratio for all clusters shown in Fig. \ref{mlratio2} is  $<\!\Upsilon_{Obs}/\Upsilon_{Kroupa}\!> = 0.97 \pm 0.03$, compatible with unity.
In particular the metal-poor clusters have $M/L$ ratios in good agreement with a Kroupa IMF. 
\citet{kruijssenmieske2009} and \citet{kimmigetal2015} found that the dynamical mass-to-light $M/L$ ratios of globular clusters are systematically lower than expected from canonical stellar population
models. This was interpreted by \citet{kruijssenmieske2009} as due to ongoing cluster dissolution. We cannot confirm their results for the majority of
globular clusters. The only clusters which are systematically below unity are the metal-rich clusters
with [Fe/H]$>-1$. This could indicate a different present-day mass function, possible due to either a different IMF or ongoing dissolution. 
However we have only five clusters with [Fe/H]$>-1$ in our sample and their $\Upsilon_{Obs}/\Upsilon_{Kroupa}$ ratios are within the range of values seen for the
low-metallicity clusters. It therefore remains an open question if the mass function of the high metallicity clusters is really different from that of the low-metallicity ones
and, if true, where this difference is coming from.

\subsection{Intermediate-mass black holes}  \label{sec:imbh}

Intermediate-mass black holes (IMBHs) are black holes in the mass range $10^2$ - $10^5$ M$_\odot$. They might provide the missing link between stellar mass
black holes formed as the end-product of stellar evolution and the supermassive black holes found in the centers of galaxies. In the last few years, evidence
for the existence of IMBHs has been accumulating. \citet{betal04} for example found a $10^5$ M$_\odot$ black hole at the center of the Seyfert 1 galaxy POX 52 
based on the broadness of the H$\beta$ profile. \citet{farrelletal2009} found evidence that the ultraluminous X-ray source in the galaxy ESO243-49
is powered by an accreting black hole with a mass between $10^2$ and $10^5$ M$_\odot$. The IMBH nature of the accreting black hole was later confirmed by \citet{webbetal2010} 
and \citet{servillatetal2011}.

Evidence for the existence of IMBHs in globular clusters is more controversial, mainly due to the fact that a centrally concentrated cluster of compact remnants
can produce a rise in the velocity dispersion profile similar to an IMBH. \citet{gerssenetal2002} found evidence for the existence of
a 4000 $M_\odot$ IMBH in the Galactic globular cluster M15 based on radial velocity measurements of individual stars near the cluster center.
However, \citet{baumgardtetal2003c} performed $N$-body simulations of star clusters without IMBHs and 
found that they could reproduce the radial velocity and surface density profile of M15 without the need for a central IMBH.
\citet{noyolaetal2010} and \citet{jalalietal2012} reported evidence for a 50000 M$_\odot$ IMBH in the globular
cluster $\omega$~Cen based on VLT-FLAMES integrated spectra of the central parts of the cluster
and detailed $N$-body models. In contrast, \citet{vdmanderson2010} found that the velocity dispersion increase in the center  
can be explained by a radially anisotropic velocity dispersion profile and derived a 1$\sigma$ upper limit of only 12000 M$_\odot$ for any possible IMBH.

\citet{lutzgendorfetal2011} presented results from ground based VLT/FLAMES spectroscopy in combination with HST data for the globular cluster NGC 6388
and found a very large central velocity dispersion of 25 km/sec in this cluster, which they could only explain by an IMBH with a mass of 
$1.7 \pm 0.9 \cdot 10^4$ M$_\odot$. \citet{lanzonietal2013} and \citet{lapennaetal2015} on the other hand obtained VLT FLAMES and KMOS spectra of 
52 and 82 giant stars near the cluster center and found a low central velocity dispersion of about 13 km/sec, which limited the mass of any central black hole
to less than 2000 M$_\odot$. In a re-analysis of all existing data, \citet{lutzgendorfetal2015} found that individual radial velocities
in the core of NGC 6388 are systematically biased towards the mean cluster velocity due to the blending of stars as a result of the high
central density. By simulating this effect using artificially created IFU data cubes, they confirmed their initial high value for the
velocity dispersion and derived an IMBH mass of $2.8 \pm 0.4 \cdot 10^4$ M$_\odot$.
IMBH detections were furthermore reported by \citet{lutzgendorfetal2013} for NGC 1904 ($M_{BH} = 3000 \pm 1000$ M$_{\odot}$) and NGC 6266 
($M_{BH} = 2000 \pm 1000$ M$_{\odot}$), 
\citet{feldmeieretal2013} for NGC 5286 ($M_{BH} = 1500 \pm 1000$ M$_{\odot}$),
\citet{ibataetal2009} for NGC 6715 ($M_{BH} \approx 9400$ M$_{\odot}$) and most recently by \citet{kamannetal2016} for NGC 6397 ($M_{BH} \approx 600$ M$_{\odot}$).

\begin{figure*}
\begin{center}
\includegraphics[width=16cm]{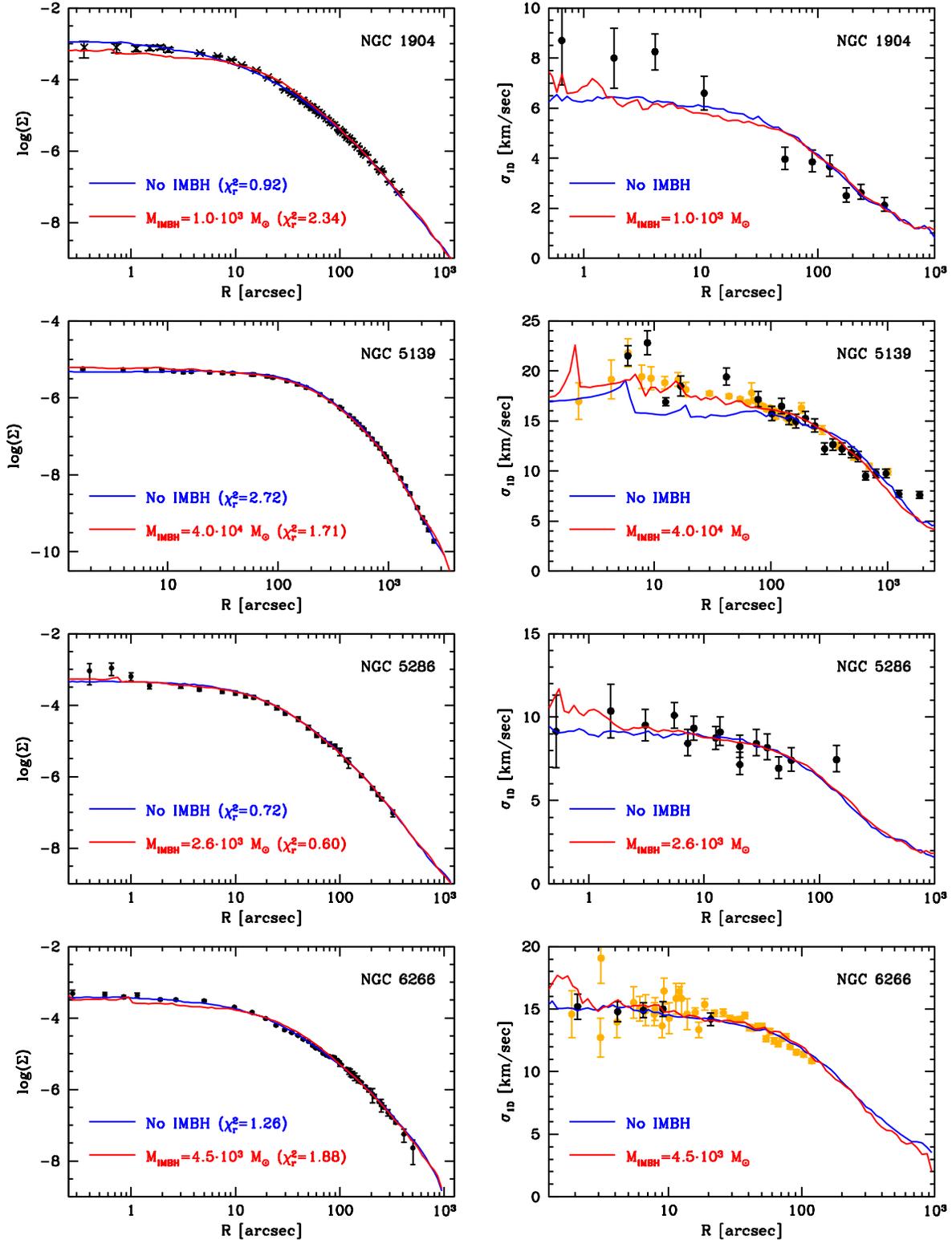}
\end{center}
\caption{Fit of the surface density profiles (left panels) and velocity dispersion profiles (right panels) of the globular clusters NGC 1904,
NGC 5139, NGC 5286 and NGC 6266 for which previous literature work found evidence for the presence of IMBHs. For each cluster we show the best-fitting $N$-body models
with (red lines) and without (blue lines) IMBHs. The IMBH models were obtained by interpolating between the grid of models containing
IMBHs between 0.5\% to 2\% of the total cluster mass. The best-fitting IMBH models fit the observed surface density profiles worse for NGC 1904 and NGC 6266
and do not improve the fits of the velocity dispersion profiles, indicating that the clusters do not contain IMBHs. For
NGC 5286 both no-IMBH and IMBH models provide a a good fit. In NGC 5139 a model with an IMBH of $4.1 \cdot 10^4$ M$_\odot$
provides a significantly better fit of the surface density and velocity dispersion profile than a no-IMBH model, making this cluster the
cluster which shows the strongest evidence for an IMBH.}
\label{figimbh1}
\end{figure*}

\begin{figure*}
\begin{center}
\includegraphics[width=16cm]{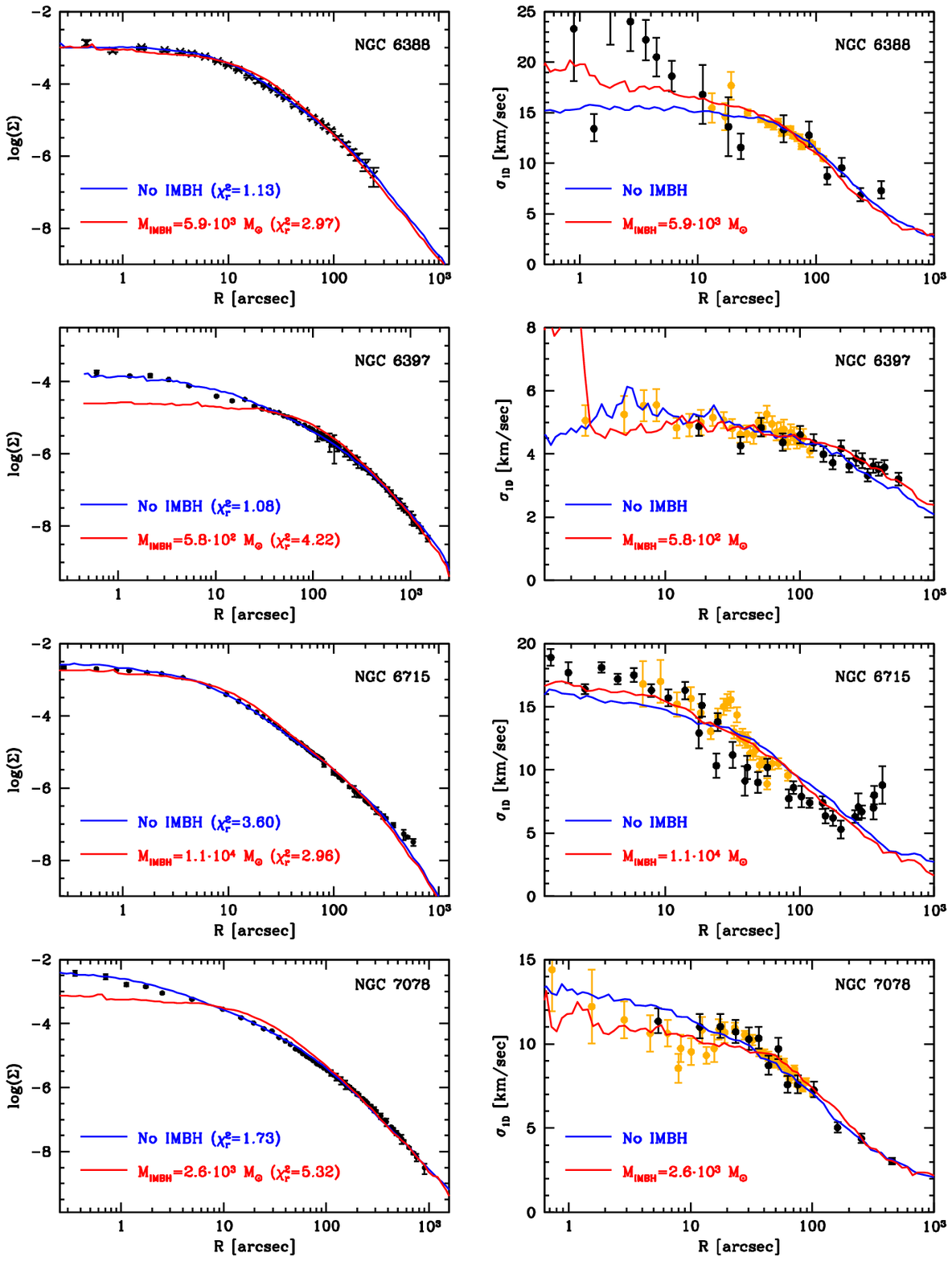}
\end{center}
\caption{Same as Fig. \ref{figimbh1} for the globular clusters NGC 6388, NGC 6397, NGC 6715 and NGC 7078. IMBHs are excluded for NGC 6397 and 7078
due to the very poor fit of the surface density profiles. In NGC 6388, IMBH models provide a less good fit to the surface density profile than the no IMBH model, however
the uncertainty about the central velocity dispersion profile prevents us from drawing any firm conclusions. In NGC 6715 a model with a $11,000$ M$_\odot$ IMBH
provides a better fit to the velocity dispersion profile than a no IMBH model, but fits the surface density profile less well.}
\label{figimbh2}
\end{figure*}

Figs. \ref{figimbh1} and \ref{figimbh2} depict the surface density profiles of the above mentioned eight clusters and compare the observed profiles with the best-fitting
no-IMBH models and the best-fitting IMBH models from our grid of $N$-body simulations. The best-fitting IMBH models were obtained by interpolating only among 
models with IMBHs. Since we calculated models containing IMBHs with masses of 0.5\%, 1\% and 2\% of the final cluster mass, the IMBH models are restricted to 
IMBH mass fractions between 0.5\% to 2\% of the cluster mass. 

In NGC 1904, the best-fitting IMBH model does significantly worse than the best-fitting no-IMBH
model, the reduced $\chi^2$ value for the IMBH model is 2.36 as opposed to 1.12 for the best-fitting IMBH model. The reason is the poor fit of the observed
surface density profile in the innermost 30 arcsec, for the velocity dispersion data alone both IMBH and no-IMBH model do about equally well. The reason for the
bad fit of the surface density profile is the fact that star clusters with IMBHs have a weak cusp in surface density as a result of mass segregation and energy equipartition
\citep{baumgardtetal2005}. This together with the fact that NGC 1904 has a relatively small relaxation time ($T \approx$ 3 Gyrs) means that no IMBH
model is able to reproduce the observed surface density profile after a Hubble time, independent of the starting density profile. We conclude that NGC 1904
does not contain an IMBH. A similar problem exists for NGC 6266 (lowest panels of Fig.  \ref{figimbh1}). The problem is even more apparent for NGC 6397 
and NGC 7078 (M15), both clusters with very steeply rising central density profiles which are in complete disagreement to how IMBH models at the same dynamical
age look like (see Fig. \ref{figimbh2}).

In NGC 5286, the IMBH model fits the observed profiles marginally better than the best-fitting no-IMBH model. However, since the best-fitting no-IMBH
model has a reduced $\chi^2$ value near one, the IMBH detection is not significant. This confirms the results of \citet{feldmeieretal2013}.
In NGC 6388 the best-fitting no-IMBH model fits the surface density profile better than the best-fitting IMBH model. Unfortunately 
the velocity dispersion profile in the central few arcsec is highly controversial in this cluster. If the central velocity dispersion is as low as 13 km/sec 
as found by \citet{lanzonietal2013} the cluster definitely 
does not contain an IMBH, while if the velocity dispersion profile is rising as found by \citet{lutzgendorfetal2015} an IMBH could be present.
Interestingly, we have difficulties reproducing both the low velocity dispersion from \citet{lanzonietal2013} with a no IMBH model as well
as the high velocity dispersion found by \citet{lutzgendorfetal2015} with our best-fitting IMBH model, which could be an indication that
both values are biased to too low/high values. A final decision on whether an IMBH is present in NGC 6388 or not can only be made once 
the velocity dispersion profile in the center of the cluster is known. 

The cluster with the strongest evidence for an IMBH is $\omega$ Cen. Our best-fitting no-IMBH model provides a very poor fit to the velocity dispersion profile. It has a reduced
$\chi_r^2$ value of 2.72, the second highest $\chi_r^2$ value of all clusters in our sample after NGC 6715. An unsegregated, isotropic star cluster without an IMBH
can therefore be safely excluded as the starting condition for $\omega$ Cen. In contrast 
an IMBH model with an IMBH of 40,000 M$_\odot$ has a reduced $\chi^2$ value of only 1.71. Fig. \ref{figimbh1} shows that this models provides a much
better fit than the no-IMBH model, the $\chi^2_r$ value is larger than one mainly because the measured data points have very small error bars of only a few hundred m/s.
Given the limited range of models which we can explore, it is difficult to reproduce any velocity dispersion profile to such a level of precision.
\citet{zocchietal2016} have argued that radially anisotropic velocity dispersion profiles could create a similar increase in the velocity dispersion
profile as a central IMBH. In addition,
\citet{vdmanderson2010} found that anisotropic models provided a better fit to the velocity dispersion profile of $\omega$ Cen than isotropic models with central IMBHs.
While the models of \citet{vdmanderson2010} took mass segregation of stars into account, it is not clear how realistic their approach was. A look at 
their Fig.~7 for example shows that
their isotropic no-IMBH model is already in very good agreement with the observed velocity dispersion profile, while it provides a very poor fit in our case.
It is therefore not clear if the inclusion of radial anisotropy would change the velocity dispersion profile by a large enough amount to bring our models 
without IMBHs into agreement with the observations,
especially since \citet{vdmanderson2010} and the proper motion data of Watkins et al. show that $\omega$ Cen is essentially isotropic in its center and
only mildly radially anisotropic beyond 200 arcsec. It seems more likely that $\omega$ Cen contains either a dark cluster of compact remnants or a population of
low-mass stars in its center on top of what mass segregation is already producing in our models, or a $\sim$ 40,000 M$_\odot$ IMBH. Comparison of the observed stellar
mass function of stars at different radii will help to further refine our models and should hopefully clarify the situation. 

In NGC 6715 a model with an IMBH of $M_{IMBH}=11,000$ M$_\odot$ provides a slightly better fit to the velocity dispersion profile of the cluster but fits the 
central surface density profile less well.  NGC 6715 (M54) has however the added complication that the cluster is the center of the Sagittarius 
dwarf galaxy so that at each radius stars that are part of the nucleus of Sagittarius contribute to the surface density and velocity dispersion profile
\citep{bellazzinietal2008}.
An increase in the fraction of Sagittarius stars is almost certainly responsible for the rise in the velocity dispersion profile seen beyond 200''. 
There is also still a significant discrepancy between the best-fitting IMBH model and the observed velocity dispersion profile of NGC 6715. Although
an IMBH might be present in NGC 6715 as well, we regard the evidence for an IMBH in NGC 6715 as weaker than in $\omega$ Cen.

\subsection{Cluster distances}

\begin{figure}
\begin{center}
\includegraphics[width=8.5cm]{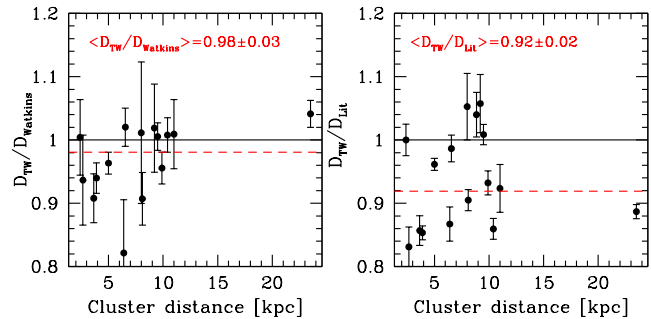}
\end{center}
\caption{Ratio of the cluster distances derived in this work to the distances found by \citet{watkinsetal2015b} (left panel) and cluster distances compiled from the
literature (right panel) for the clusters for which we have determined distances. Dashed lines show the average distance ratio in each panel. Our distances agree
very well with the distances of \citet{watkinsetal2015b}, but are on average about 8\% smaller than the literature distances.}
\label{distance}
\end{figure}

Fig.~\ref{distance} compares the cluster distances derived in this work with the distances derived by \citet{watkinsetal2015b} (left panel) and cluster distances from the literature
(right panel). In order to derive the distances, we fitted our models to the surface density, radial velocity and proper motion dispersion profiles of the clusters
and varied the distance until the combined $\chi^2$ was minimal. Our distances are on average 2\% $\pm$ 3\% smaller than those of \citet{watkinsetal2015b} and
hence in excellent agreement with their distances. 
The discrepancy is larger when we compare with literature distances which are mainly obtained from CMD fitting since our
distances are on average 8\% smaller. The literature distances were taken mostly from  \citet{ferraroetal1999}, however we obtain a similar difference when
using the \citet{harris1996} distances. We find no obvious correlation between the distance ratio $D_{TW}/D_{Lit}$ and any other cluster parameter like metallicity,
total mass or cluster distance (see Fig.~\ref{distance}). It therefore remains unclear where the discrepancy between our distances and the literature values is coming from.
Parallax data from the {\it GAIA} satellite should help to settle the globular cluster distance scale.

\subsection{Deviations from Newtonian dynamics?}

\citet{scarpaetal2007b} and \citet{scarpaetal2011} reported evidence for a flattening of the velocity dispersion profile in the outermost parts of
a number of globular clusters including NGC 288, NGC 1851, NGC 1904 and NGC 7099, which they attributed to a deviation from Newtonian dynamics.
Our models give us a chance to verify their claims. As can be seen from Figs.~\ref{fig1a}, \ref{fig2a} and \ref{fig19a}, the measured velocity dispersion 
profiles of the four clusters studied by Scarpa et al. are compatible with predictions of our $N$-body models out to the
outermost data points with no evidence for a breakdown of Newtonian mechanics. The same is the case for most other clusters. We attribute the difference
to Scarpa et al. to the fact that we calculate velocity dispersion
profiles based on a larger number of radial velocities, which allows us to more efficiently identify binaries and non-members. In addition we apply a $\chi^2$ test 
based on the local velocity dispersion to separate members from non-members, while Scarpa et al. include all stars as members that have radial velocities within
certain velocity limits. In the outer parts of globular clusters, where a larger fraction of stars are non-members, the approach used by Scarpa et al. is likely to overestimate 
the velocity dispersion. The agreement with our models could probably be improved further since the simulations presented here do not include tidal fields, which 
increase the velocity dispersion of stars near the tidal radius \citep{kuepperetal2010,claydon2015}.

The only cluster which deviates significantly from our predictions is NGC~6715 (M54), where the velocity dispersion profile rises in the outermost
few 100''. As discussed by \citet{bellazzinietal2008}, this is most likely due to the fact that the sample includes stars from the Sagittarius dwarf galaxy, which follow
a different kinematical profile and whose relative contribution increases in the outermost cluster parts. We have therefore neglected all data points beyond
200'' in the dynamical analysis of this cluster. Apart from NGC~6715, the globular cluster velocity dispersion
profiles do not show any evidence for deviations from Newtonian dynamics out to distances of several 100 arcsec, corresponding to a physical
distance of $\sim$10~pc.

\section{Conclusions and Outlook}

We have run a large grid of 900 $N$-body simulations of star clusters, varying the initial half-mass radius, density profile, cluster metallicity, and the mass fraction 
of a central IMBH. We have also determined new radial velocity dispersion profiles of 50 Galactic globular clusters from about 25,000 published line-of-sight 
radial velocity measurements of stars in globular clusters, and combined these profiles with velocity dispersion data based on proper motions and published surface density profiles. 
By comparing the $N$-body data with the observed data and selecting the best-fitting model for each cluster, we were then able to derive absolute masses, mass-to-light ratios 
and limits on the possible presence of IMBH in the centers of all clusters. For a subset of 16 clusters for which both good proper motion and radial velocity
information is available we also determined the cluster distances.

We find that the average mass-to-light ratio
of Galactic globular clusters is $<\!\!M/L_V\!\!>=1.98 \pm 0.03$ which agrees very well with the expected $M/L_V$ ratio for stars that formed with
a standard Kroupa or Chabrier initial mass function. The mass-to-light ratios of high metallicity clusters
with [Fe/H]$>-1$ could be slightly lower then predicted by standard stellar mass functions. The number of high-metallicity clusters in our sample is however small
and the variation seen for them is within the variation found for low-metallicity clusters.
Given the good agreement between the derived and the theoretically expected $M/L$ ratios, there is
no evidence that globular cluster $M/L$ ratios are significantly effected by ongoing cluster dissolution. More accurate $M/L$ ratios, or $M/L$ ratios
for a wider range of cluster parameters will be necessary to determine what role dissolution has played for globular clusters.

We find strong evidence that $\omega$ Cen hosts an intermediate-mass black hole (IMBH) of $\sim 40,000$ M$_\odot$ in its center since the velocity dispersion profile of the
cluster is in strong disagreement to $N$-body models without an IMBH. A compact cluster of stellar remnants
in the center or a cluster that starts with a radially anisotropic velocity dispersion profile
might be alternatives to an IMBH, however these possibilities seem unlikely given how well our isotropic, non mass-segregated models fit all other clusters.
Given the absence of radio and X-ray emission from the center of $\omega$~Cen \citep{mft05,haggardetal2013}, this result implies that if an IMBH exists in the center,
it must accrete very little or with very low efficiency ($\eta<10^{-9}$). 
Evidence for the presence of an
IMBH is also found in NGC 6715 (M54), however in NGC 6715 the best-fitting IMBH model is still in significant disagreement with the velocity
dispersion profile.
We can strongly exclude the presence of IMBHs in NGC 6397 and M15 and find that they are also unlikely to be present in most other clusters since IMBH models
provide significantly less good fits to the surface density profiles than no-IMBH models. We therefore conclude that if IMBHs exist in globular clusters, 
they can only exist in a small fraction of them. 

In the present work we only compared the observed velocity dispersion and surface density profiles with results from our $N$-body simulations. The next
step is to also compare the mass function of stars at different radii with our predictions by performing simulations of star clusters which start with a range
of initial mass functions and performing simulations that include cluster dissolution due to external tidal fields. Mass functions of stars have been observed for about 
half of all globular clusters from our sample by \citet{demarchietal2007,paustetal2010}, and \citet{sollimaetal2012}. Comparison of the stellar mass functions will allow to 
accurately predict the structural parameter of the clusters like core and half-mass radii and the corresponding densities and relaxation times. It will also
allow to determine the starting conditions of globular clusters in terms of initial radii, initial mass functions and the amount of primordial mass segregation 
and thereby gain a much better understanding of their formation and evolution.

\section*{Acknowledgments}

We thank Karl Gebhardt, Georges Meylan, Ben MacLean and Sandro Villanova for sharing unpublished data with us. We are also grateful to Long Wang
for sending us data from his simulations. We finally thank Pouria Khalaj and two anonymous referees for
comments that helped improve the presentation of the paper. This work has made use of BaSTI web tools.

\bibliographystyle{mn2e}
\bibliography{mybib}

\section*{Appendix A: Sources for the photometric and spectroscopic information of individual clusters}

\begin{table}
\caption{Sources for velocity and surface density data used in this work (LOS = line-of-sight radial velocities,
  PM = proper motion velocity dispersion profile, IFU = Integral field unit velocity dispersion, SD = surface density profile)}
\begin{tabular}[h]{c@{\hspace{0.2cm}}c@{\hspace{0.2cm}}c}
\hline
Name & Source & Type \\ 
\hline
NGC 104  & \citet{mayoretal1983} & LOS \\
         & \citet{gebhardtetal1995}& LOS \\
         & \citet{mclaughlinetal2006} & PM \\
         & \citet{carrettaetal2009} & LOS\\
         & \citet{laneetal2011} & LOS \\
         & \citet{grattonetal2013} & LOS \\
         & \citet{watkinsetal2015a} & PM \\
         & \citet{dacosta2016} & LOS \\
         & \citet{marinoetal2016} & LOS \\[+0.1cm]
NGC 288  & \citet{pryoretal1991} & LOS \\
         & \citet{laneetal2011} & LOS \\
         & \citet{lucatelloetal2015} & LOS \\
         & \citet{watkinsetal2015a} & PM \\
         & \citet{dacosta2016} & LOS \\[+0.1cm]
NGC 362  & \citet{fischeretal1993} & LOS \\ 
         & \citet{carrettaetal2013} & LOS \\
         & \citet{dorazietal2015} & LOS \\
         & \citet{schoenebeck2015} & LOS \\
         & \citet{watkinsetal2015a} & PM \\[+0.1cm]
NGC 1851 & \citet{scarpaetal2011} & LOS \\
         & \citet{grattonetal2012} & LOS \\
	 & \citet{lutzgendorfetal2013} & IFU \\
	 & \citet{lardoetal2015}  & LOS \\
         & \citet{lucatelloetal2015} & LOS \\
         & \citet{watkinsetal2015a} & PM, LOS \\[+0.1cm]
NGC 1904 & \citet{carrettaetal2009} & LOS\\
         & \citet{scarpaetal2011} & LOS \\
         & \citet{lutzgendorfetal2013} & IFU \\
         & \citet{dorazietal2015} & LOS \\[+0.1cm]
NGC 2419 & \citet{baumgardtetal2009} & LOS \\
         & \citet{ibataetal2011} & LOS \\
         & \citet{bellazzini2007} & SD \\[+0.1cm]
NGC 2808 & \citet{cacciarietal2004} & LOS \\
         & \citet{carrettaetal2006} & LOS \\
         & \citet{grattonetal2011} & LOS \\
         & \citet{lutzgendorfetal2012} & IFU, LOS \\
         & \citet{marinoetal2014} & LOS \\
         & \citet{lardoetal2015} & LOS \\
         & \citet{dorazietal2015} & LOS \\
         & \citet{wangetal2016b} & LOS \\
         & \citet{watkinsetal2015a} & PM, LOS \\[+0.1cm]
NGC 3201 & \citet{coteetal1995} & LOS \\
         & \citet{carrettaetal2009} & LOS \\
         & \citet{mucciarelli2015} & LOS\\[+0.1cm]
NGC 4147 & \citet{kimmigetal2015} & LOS \\
         & \citet{villanovaetal2016b} & LOS \\[+0.1cm]
NGC 4372 & \citet{kacharovetal2014} & LOS, SD \\
         & \citet{lardoetal2015} & LOS \\[+0.1cm]
NGC 4590 & \citet{carrettaetal2009} & LOS \\
         & \citet{laneetal2011}& LOS \\[+0.1cm]
NGC 4833 & \citet{carrettaetal2014} & LOS \\
         & \citet{lardoetal2015} & LOS \\
         & \citet{melbourneetal2000} & SD \\[+0.1cm]
NGC 5024 & \citet{laneetal2011} & LOS \\
         & \citet{kimmigetal2015} & LOS \\[+0.1cm]
NGC 5053 & \citet{bobergetal2015} & LOS \\
         & \citet{kimmigetal2015} & LOS \\[+0.1cm]
\hline
\end{tabular}
\end{table}

\setcounter{table}{2}

\begin{table}
\caption{continued}
\begin{tabular}[h]{c@{\hspace{0.2cm}}c@{\hspace{0.2cm}}c}
\hline
Name & Source & Type \\
\hline
NGC 5139 & \citet{mayoretal1997} & LOS \\
         & \citet{reijnsetal2006} & LOS \\
         & \citet{vandevenetal2006} & PM \\
         & \citet{pancinoetal2007} & LOS \\
         & \citet{johnsonetal2008} & LOS \\
         & \citet{sollimaetal2009} & LOS \\
         & \citet{noyolaetal2010} & IFU \\
         & \citet{scarpaetal2010} & LOS \\
         & \citet{dacosta2012} & LOS \\
         & \citet{villanovaetal2014}& LOS \\
         & \citet{watkinsetal2015a} & PM \\
         & \citet{gebhardtetal2016}& LOS \\[+0.1cm]
NGC 5272 & \citet{gg1979} & LOS \\
         & \citet{pilaetal2000} & LOS \\
         & \citet{smolinskietal2011} & LOS \\
         & \citet{kamannetal2014} & LOS \\
         & \citet{kimmigetal2015} & LOS \\[+0.1cm]
NGC 5286 & \citet{feldmeieretal2013} & IFU, LOS, SD\\
         & \citet{marinoetal2015} & LOS\\[+0.1cm]
NGC 5466 & \citet{pryoretal1991} & LOS \\
         & \citet{kimmigetal2015} & LOS \\[+0.1cm]
NGC 5694 & \citet{lutzgendorfetal2013} & IFU, SD \\
         & \citet{bellazzinietal2015} & LOS \\[+0.1cm]
NGC 5824 & \citet{lutzgendorfetal2013} & IFU, SD \\
         & \citet{roedereretal2016} & LOS \\[+0.1cm]
NGC 5904 & \citet{rastorguevsamus1991} & LOS \\
         & \citet{battagliaetal2008} & LOS \\
         & \citet{carrettaetal2009} & LOS \\
         & \citet{grattonetal2013} & LOS \\
         & \citet{kimmigetal2015} & LOS \\
         & \citet{watkinsetal2015a} & PM \\[+0.1cm]
NGC 5927 & \citet{simmereretal2013} & LOS \\
         & \citet{lardoetal2015} & LOS \\
         & \citet{watkinsetal2015a} & PM \\[+0.1cm]
NGC 6093 & \citet{lutzgendorfetal2013} & IFU, LOS, SD \\
         & \citet{carrettaetal2015} & LOS\\[+0.1cm]
NGC 6121 & \citet{petersonetal1995} & LOS \\
         & \citet{ivansetal1999} & LOS \\
         & \citet{carrettaetal2009} & LOS \\
         & \citet{laneetal2011}& LOS \\
         & \citet{malavoltaetal2015} & LOS \\
         & \citet{macleanetal2016} & LOS \\[+0.1cm]
NGC 6139 & \citet{bragagliaetal2015} & LOS \\[+0.1cm]
NGC 6171 & \citet{piateketal1994} & LOS \\
         & \citet{scarpaetal2007b} & LOS \\
         & \citet{carrettaetal2009} & LOS \\[+0.1cm]
NGC 6205 & \citet{luptonetal1987} & LOS \\
         & \citet{pilaetal2000} & LOS \\
         & \citet{leeetal2008} & LOS \\
         & \citet{meszarosetal2009} & LOS \\
         & \citet{smolinskietal2011} & LOS \\
         & \citet{kamannetal2014} & LOS \\
NGC 6218 & \citet{carrettaetal2007b} & LOS \\
         & \citet{laneetal2011}& LOS \\[+0.1cm]
NGC 6254 & \citet{carrettaetal2009} & LOS \\[+0.1cm]
NGC 6266 & \citet{mcnamaraetal2012} & PM \\
         & \citet{lutzgendorfetal2013} & IFU \\
         & \citet{watkinsetal2015a} & PM \\[+0.1cm]
NGC 6273 & \citet{johnsonetal2015} & LOS \\ 
         & \citet{yongetal2016} & LOS \\[+0.1cm]
\hline
\end{tabular}
\end{table}

\setcounter{table}{2}

\begin{table}
\caption{continued}
\begin{tabular}[h]{c@{\hspace{0.2cm}}c@{\hspace{0.2cm}}c}
\hline
Name & Source & Type \\
\hline
NGC 6341 & \citet{pilaetal2000} & LOS \\
         & \citet{drukieretal2007} & LOS \\
         & \citet{meszarosetal2009} & LOS \\
         & \citet{kamannetal2014} & LOS \\
         & \citet{kimmigetal2015} & LOS \\
         & \citet{watkinsetal2015a} & PM \\[+0.1cm]
NGC 6362 & \citet{watkinsetal2015a} & PM \\[+0.1cm]
NGC 6388 & \citet{carrettaetal2009} & LOS \\
         & \citet{lanzonietal2013} & LOS \\
         & \citet{lutzgendorfetal2013} & IFU, SD \\
         & \citet{lapennaetal2015} & LOS \\
         & \citet{watkinsetal2015a} & PM \\[+0.1cm]
NGC 6397 & \citet{meylanmayor1991} & LOS \\
         & \citet{gebhardtetal1995} & LOS \\
         & \citet{miloneetal2006} & LOS \\
         & \citet{carrettaetal2009} & LOS \\
         & \citet{lindetal2009} & LOS \\
         & \citet{lovisietal2012} & LOS \\
         & \citet{kamannetal2016} & LOS \\
         & \citet{watkinsetal2015a} & PM \\[+0.1cm]
NGC 6402 & \citet{kimmigetal2015} & LOS \\[+0.1cm]
NGC 6441 & \citet{watkinsetal2015a} & PM \\[+0.1cm]
NGC 6535 & \citet{zaritskyetal2014} & LOS \\
         & \citet{watkinsetal2015a} & PM \\[+0.1cm]
NGC 6624 & \citet{watkinsetal2015a} & PM \\[+0.1cm]
NGC 6656 & \citet{petersonetal1994} & LOS \\
         & \citet{coteetal1996}& LOS \\
         & \citet{laneetal2011}& LOS \\
         & \citet{grattonetal2014}& LOS \\
         & \citet{watkinsetal2015a} & PM \\[+0.1cm]
NGC 6681 & \citet{watkinsetal2015a} & PM \\[+0.1cm]
NGC 6715 & \citet{bellazzinietal2008} & LOS \\
         & \citet{ibataetal2009} & LOS \\
         & \citet{carrettaetal2010a} & LOS \\
         & \citet{watkinsetal2015a} & PM \\[+0.1cm]
NGC 6723 & \citet{rojasetal2016}  & LOS \\
         & \citet{grattonetal2015}  & LOS \\[+0.1cm]
NGC 6752 & \citet{carrettaetal2007a} & LOS \\
         & \citet{laneetal2011}& LOS \\
         & \citet{lovisietal2013}& LOS \\
         & \citet{lardoetal2015} & LOS \\
         & \citet{watkinsetal2015a} & PM \\[+0.1cm]
NGC 6809 & \citet{carrettaetal2009} & LOS \\
         & \citet{laneetal2011}& LOS\\[+0.1cm]
NGC 6838 & \citet{pl1986} & LOS \\
         & \citet{carrettaetal2009} & LOS \\
         & \citet{smolinskietal2011} & LOS \\
         & \citet{kimmigetal2015} & LOS \\
         & \citet{corderoetal2015} & LOS \\
         & \citet{drukieretal1992} & SD \\[+0.1cm]
NGC 7078 & \citet{petersonetal1989} & LOS \\
         & \citet{drukieretal1998} & LOS \\
         & \citet{gebhardtetal2000} & LOS \\
         & \citet{gerssenetal2002} & LOS \\
         & \citet{carrettaetal2009} & LOS \\
         & \citet{lardoetal2015} & LOS \\
         & \citet{kimmigetal2015} & LOS \\
         & \citet{mcnamaraetal2003} & PM \\
         & \citet{watkinsetal2015a} & PM \\[+0.1cm]
\hline
\end{tabular}
\end{table}

\setcounter{table}{2}

\begin{table}
\caption{continued}
\begin{tabular}[h]{c@{\hspace{1.5cm}}c@{\hspace{1.5cm}}c}
\hline
Name & Source & Type \\
\hline
NGC 7089 & \citet{pryoretal1986} & LOS \\
         & \citet{leeetal2008} & LOS \\
         & \citet{schoenebeck2015} & LOS \\
         & \citet{kimmigetal2015} & LOS \\[+0.1cm]
NGC 7099 & \citet{gebhardtetal1995} & LOS \\
         & \citet{scarpaetal2007a} & LOS \\
         & \citet{carrettaetal2009} & LOS \\
         & \citet{laneetal2011}& LOS \\[+0.1cm]
Terzan 8 & \citet{carrettaetal2014b} & LOS \\ 
         & \citet{salinasetal2012} & SD \\ 
\hline
\end{tabular}
\label{sourcestab}
\end{table}

\section*{Appendix B: Fits of the surface density and velocity dispersion profiles of individual clusters}

\begin{figure*}
\begin{center}
\includegraphics[width=17cm]{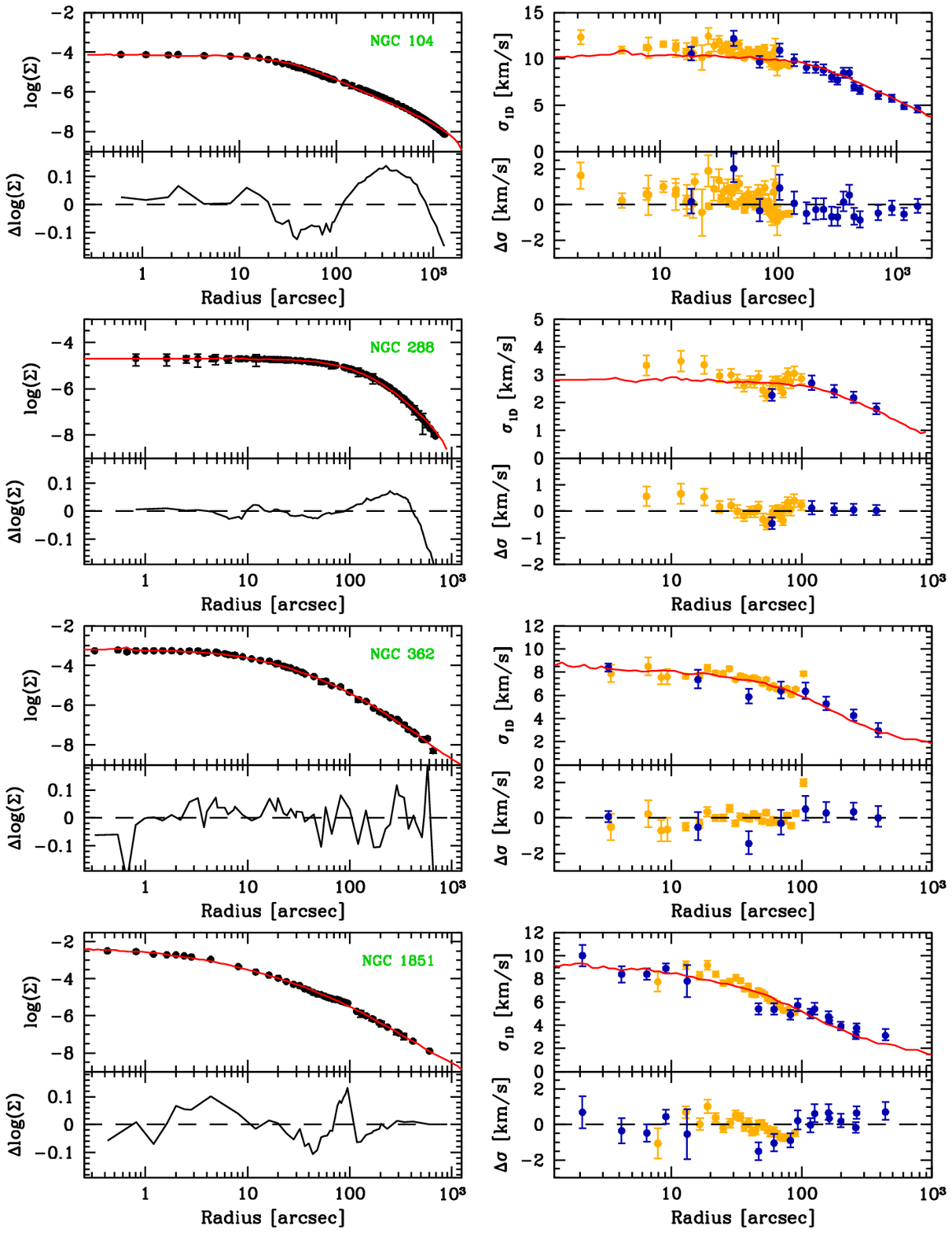}
\end{center}
\vspace*{-0.5cm}
\caption{Fit of the surface density profiles (left panels) and velocity dispersion profiles (right panels) for NGC 104,
NGC 288, NGC 362 and NGC 1851. The surface densities in the left panels are normalised to 1. In the right panels, the observed proper motion velocity dispersion profile
is shown by orange circles while the radial velocity dispersion profile derived in this work is shown 
by blue circles. Red curves show the surface density (left panel) and line-of-sight velocity dispersion (right panel) of the best-fitting $N$-body model without an IMBH 
for each cluster. The $N$-body data provides an excellent fit to the observed data for the depicted clusters. The lower panels show the differences between
the observed data and the $N$-body models.}
\label{fig1a}
\end{figure*}

\begin{figure*}
\begin{center}
\includegraphics[width=17cm]{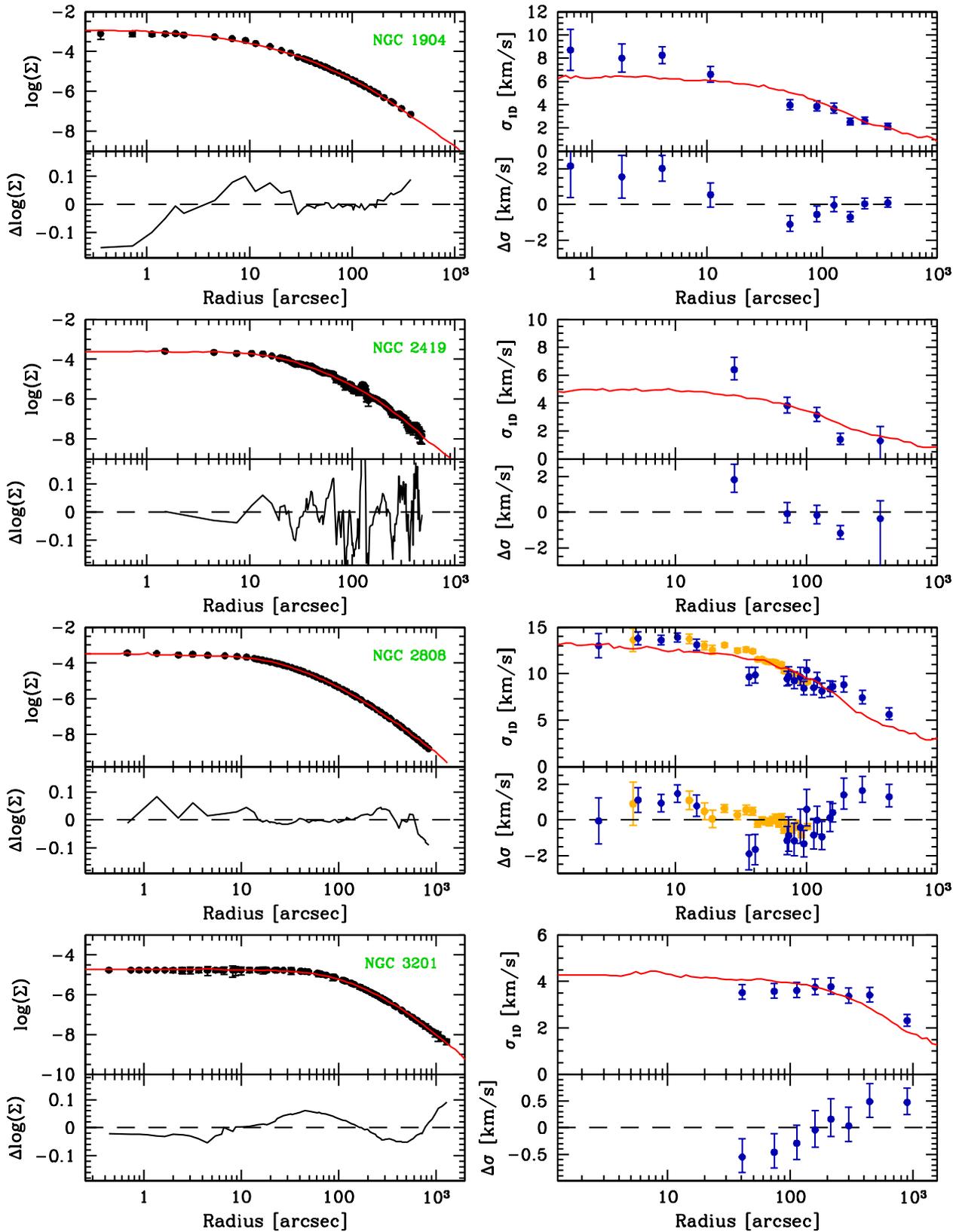}
\end{center}
\caption{Same as Fig. \ref{fig1a} for NGC 1904, NGC 2419, NGC 2808 and NGC 3201.\hspace*{7cm}}
\label{fig2a}
\end{figure*}

Figs. \ref{fig1a} to \ref{fig13a} depict our fits of the observed surface density and velocity dispersion profiles for all studied clusters.
The surface densities in the left panels are normalised to~1. In the right panels, the proper motion data 
are shown by orange circles while the radial velocity dispersion profiles derived in this work are shown by blue circles.
The predictions of the best-fitting $N$-body models are shown as solid, red lines. For clarity we show only the radial velocity dispersion 
profiles. The proper motion velocity dispersion profiles are only a few \% higher due to mass segregation.
The $N$-body models shown are the best-fitting no-IMBH models except for NGC 5139 and NGC 6715, which
show the best-fitting IMBH models.

\newpage

\begin{figure*}
\begin{center}
\includegraphics[width=17cm]{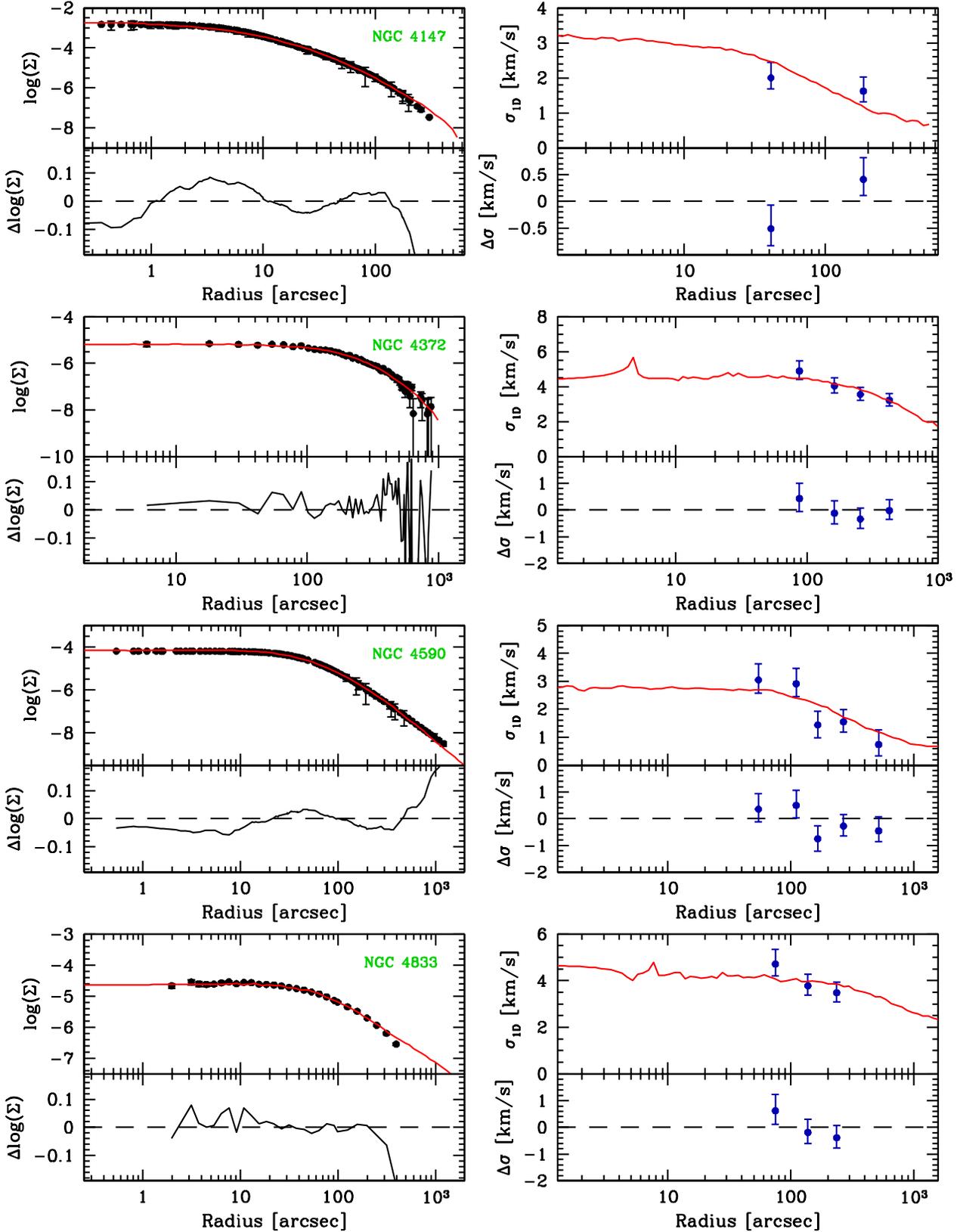}
\end{center}
\caption{Same as Fig. \ref{fig1a} for NGC 4147, NGC 4372, NGC 4590 and NGC 4833.\hspace*{7cm}}
\end{figure*}

\begin{figure*}
\begin{center}
\includegraphics[width=17cm]{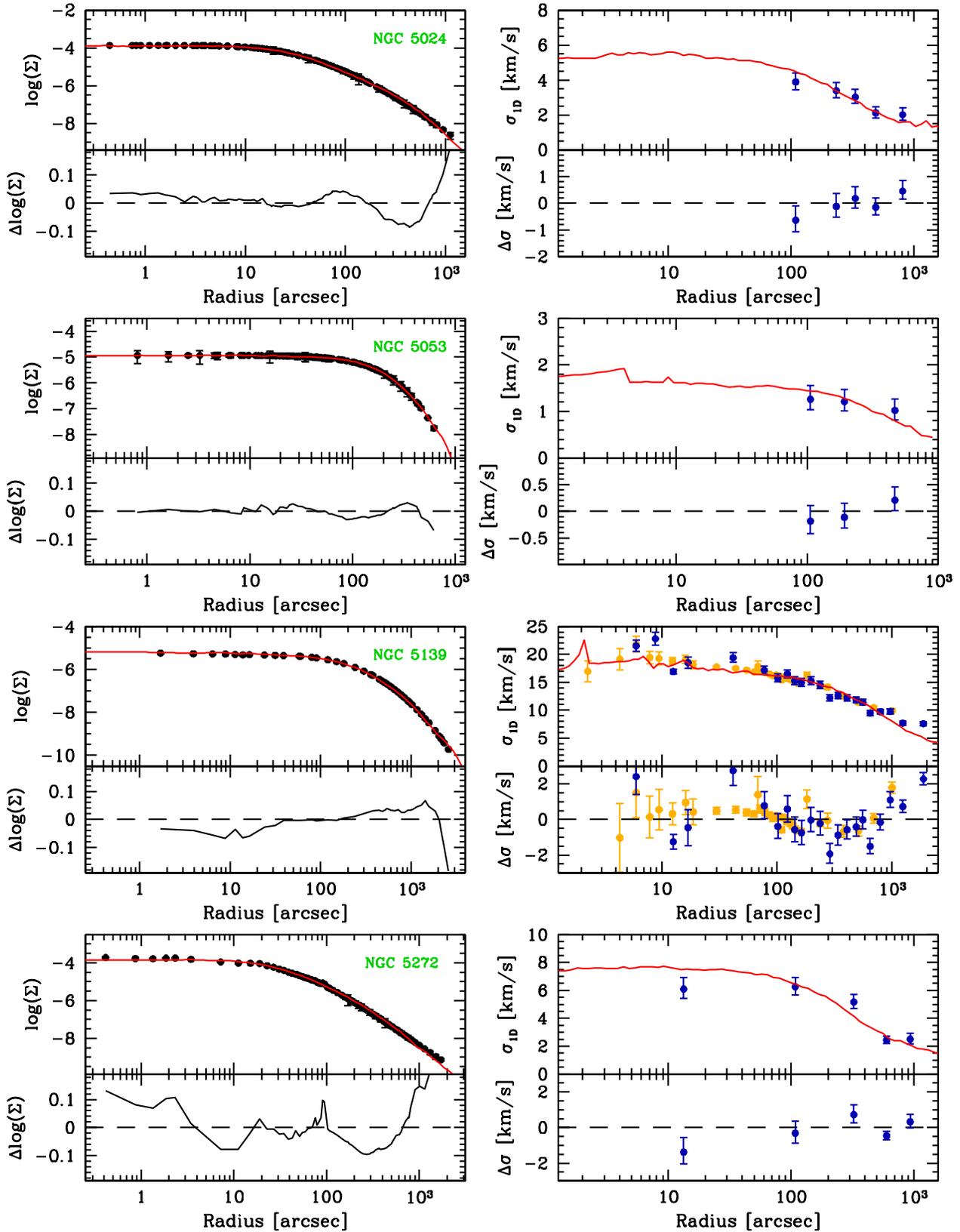}
\end{center}
\caption{Same as Fig. \ref{fig1a} for NGC 5024, NGC 5053, NGC 5139 and NGC 5272. The red, solid lines for NGC 5139 show the best-fitting
IMBH model.\hspace*{2cm}}
\end{figure*}

\begin{figure*}
\begin{center}
\includegraphics[width=17cm]{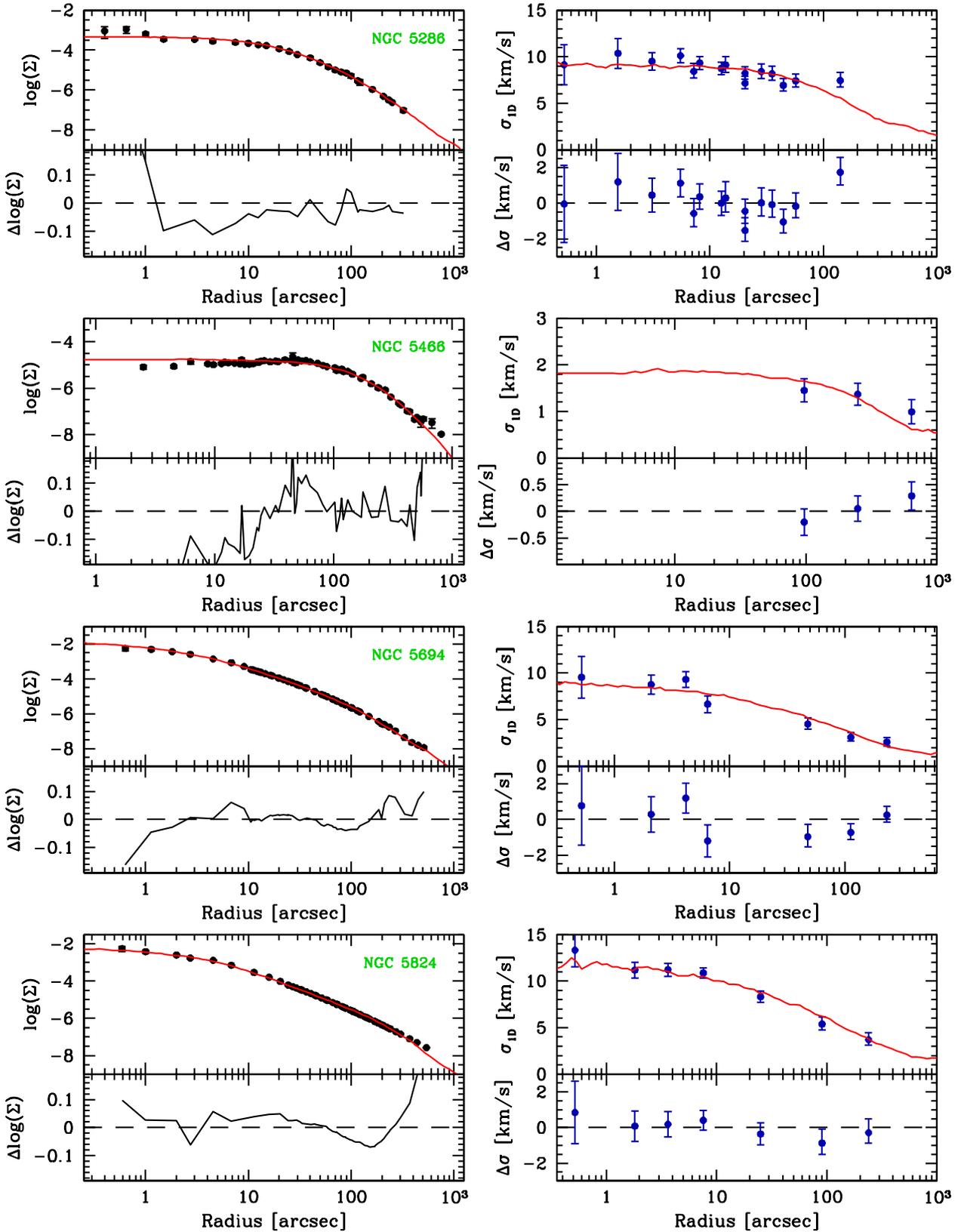}
\end{center}
\caption{Same as Fig. \ref{fig1a} for NGC 5286, NGC 5466, NGC 5694 and NGC 5824.\hspace*{7cm}}
\end{figure*}

\begin{figure*}
\begin{center}
\includegraphics[width=17cm]{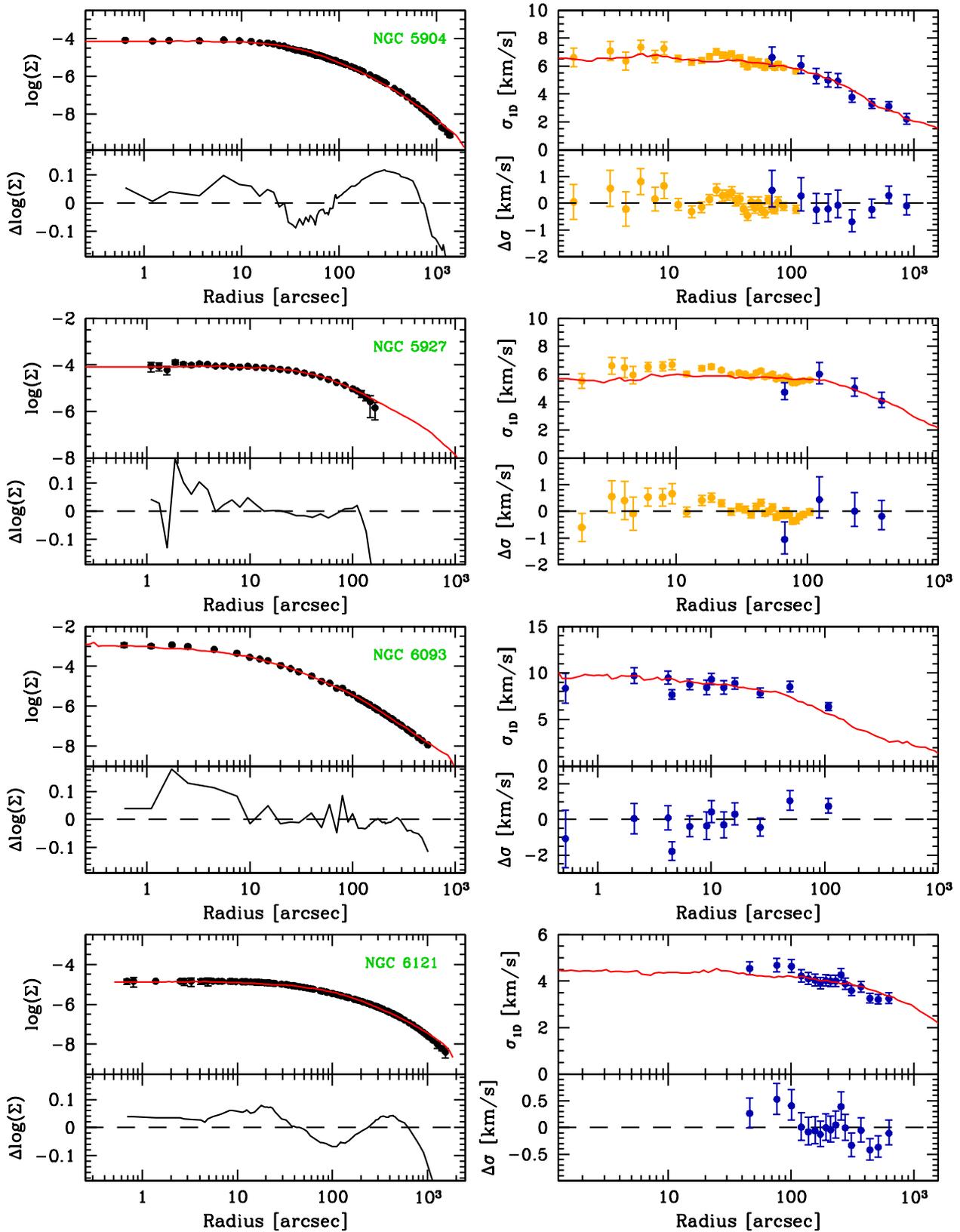}
\end{center}
\caption{Same as Fig. \ref{fig1a} for NGC 5904, NGC 5927, NGC 6093 and NGC 6121.\hspace*{7cm}}
\end{figure*}

\begin{figure*}
\begin{center}
\includegraphics[width=17cm]{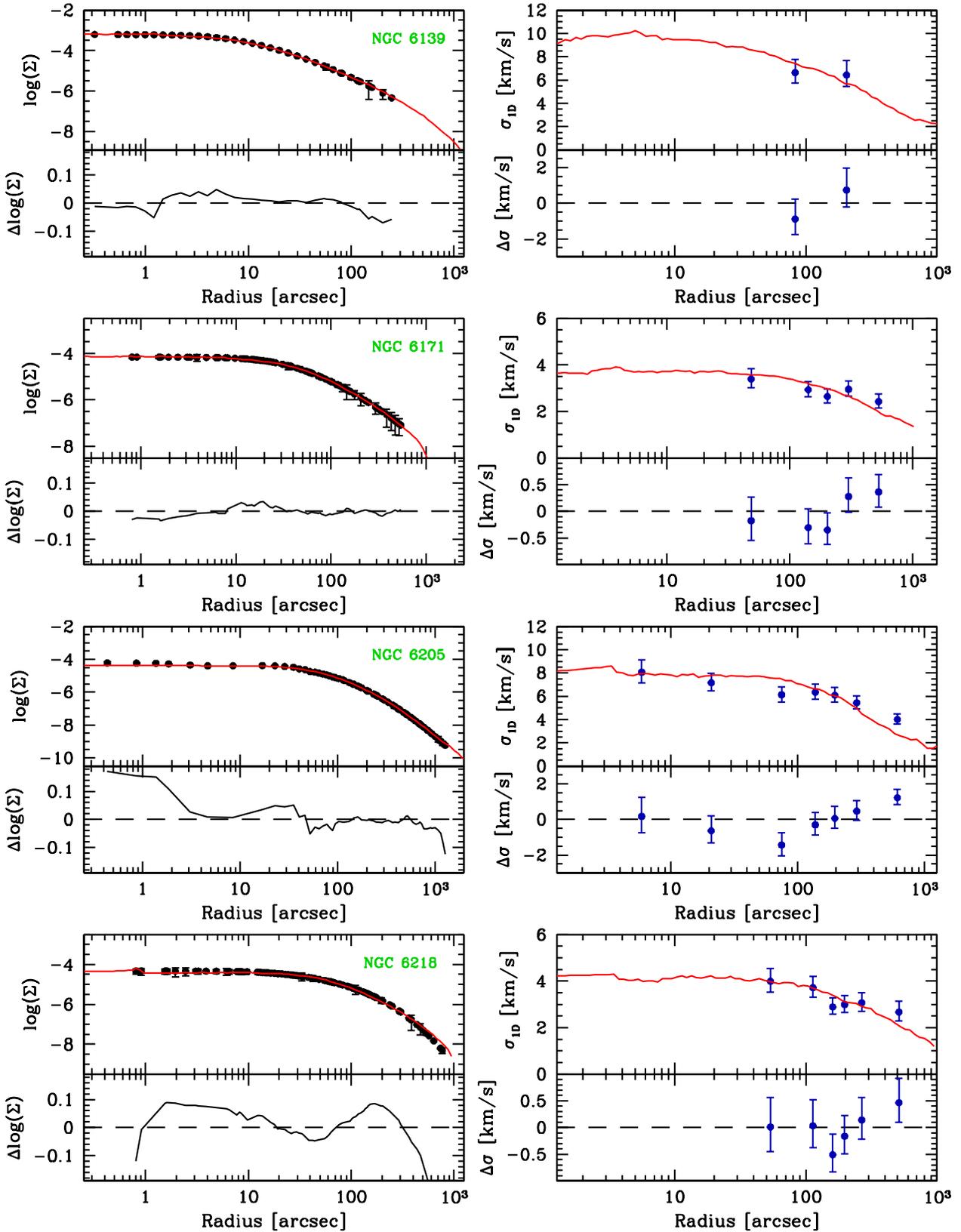}
\end{center}
\caption{Same as Fig. \ref{fig1a} for NGC6139, NGC 6171, NGC 6205, and NGC 6218.\hspace*{7cm}}
\end{figure*}

\begin{figure*}
\begin{center}
\includegraphics[width=17cm]{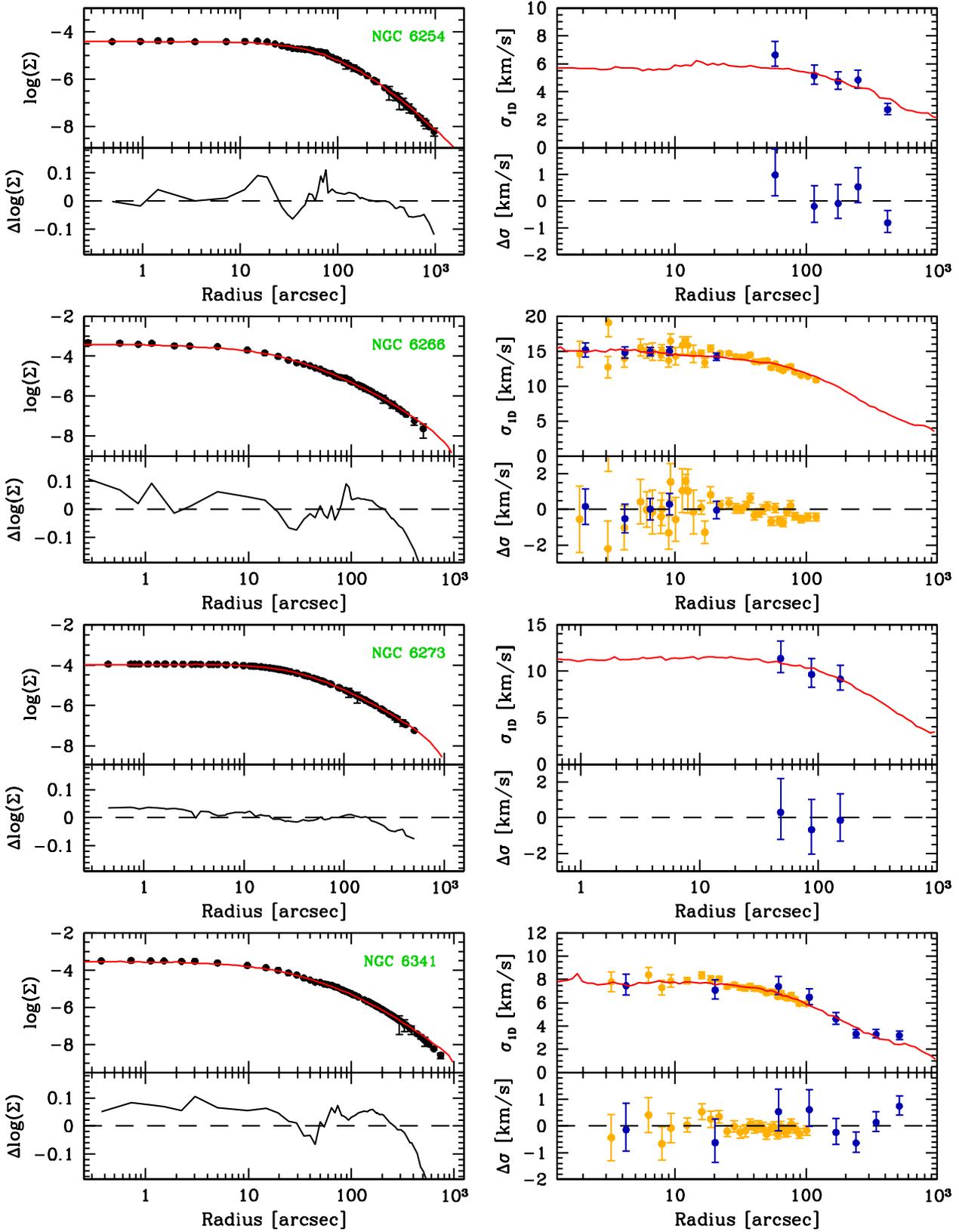}
\end{center}
\caption{Same as Fig. \ref{fig1a} for NGC 6254, NGC 6266, NGC 6273 and NGC 6341.\hspace*{7cm}}
\end{figure*}

\begin{figure*}
\begin{center}
\includegraphics[width=17cm]{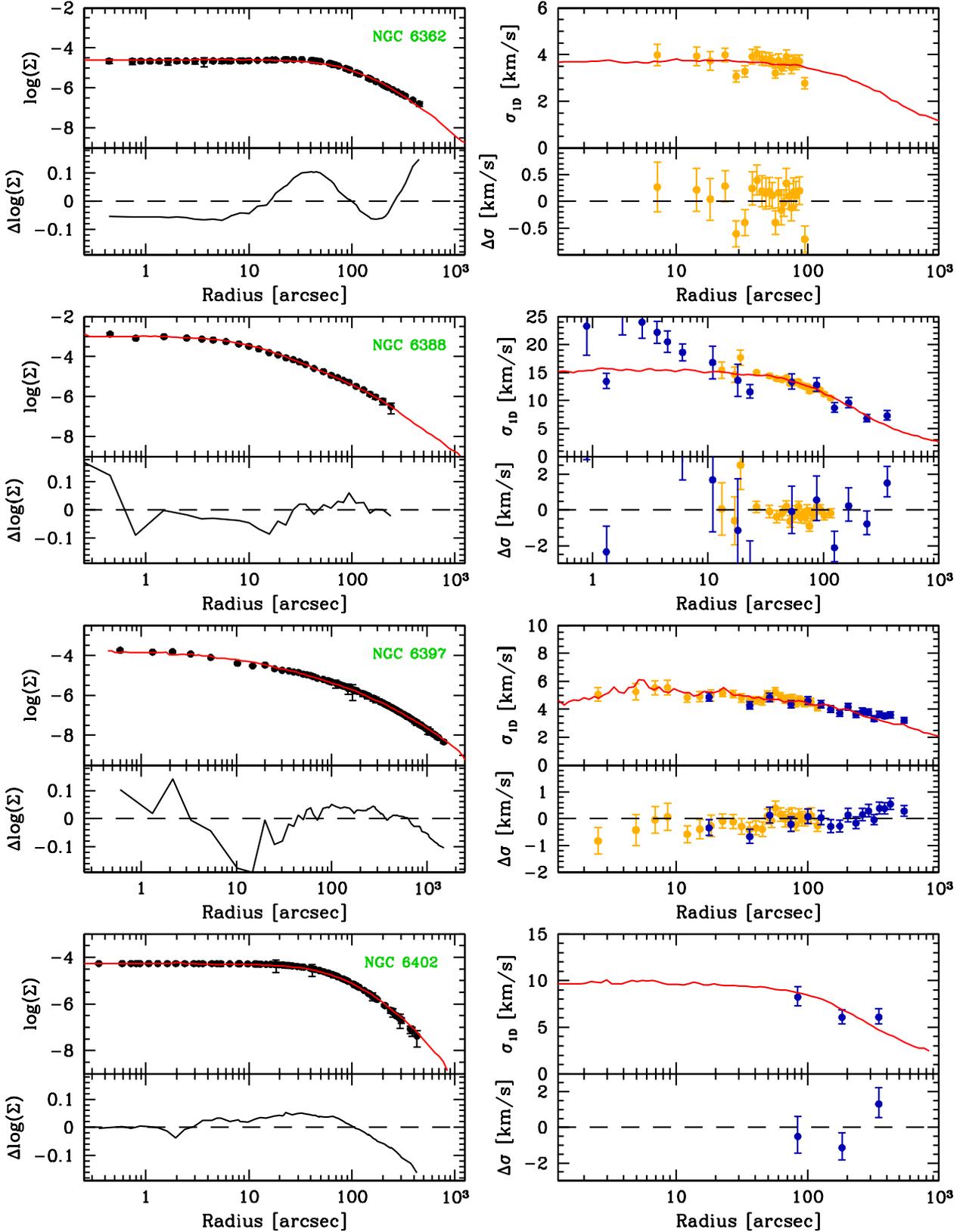}
\end{center}
\caption{Same as Fig. \ref{fig1a} for NGC 6362, NGC 6388, NGC 6397 and NGC 6402.\hspace*{7cm}}
\end{figure*}

\begin{figure*}
\begin{center}
\includegraphics[width=17cm]{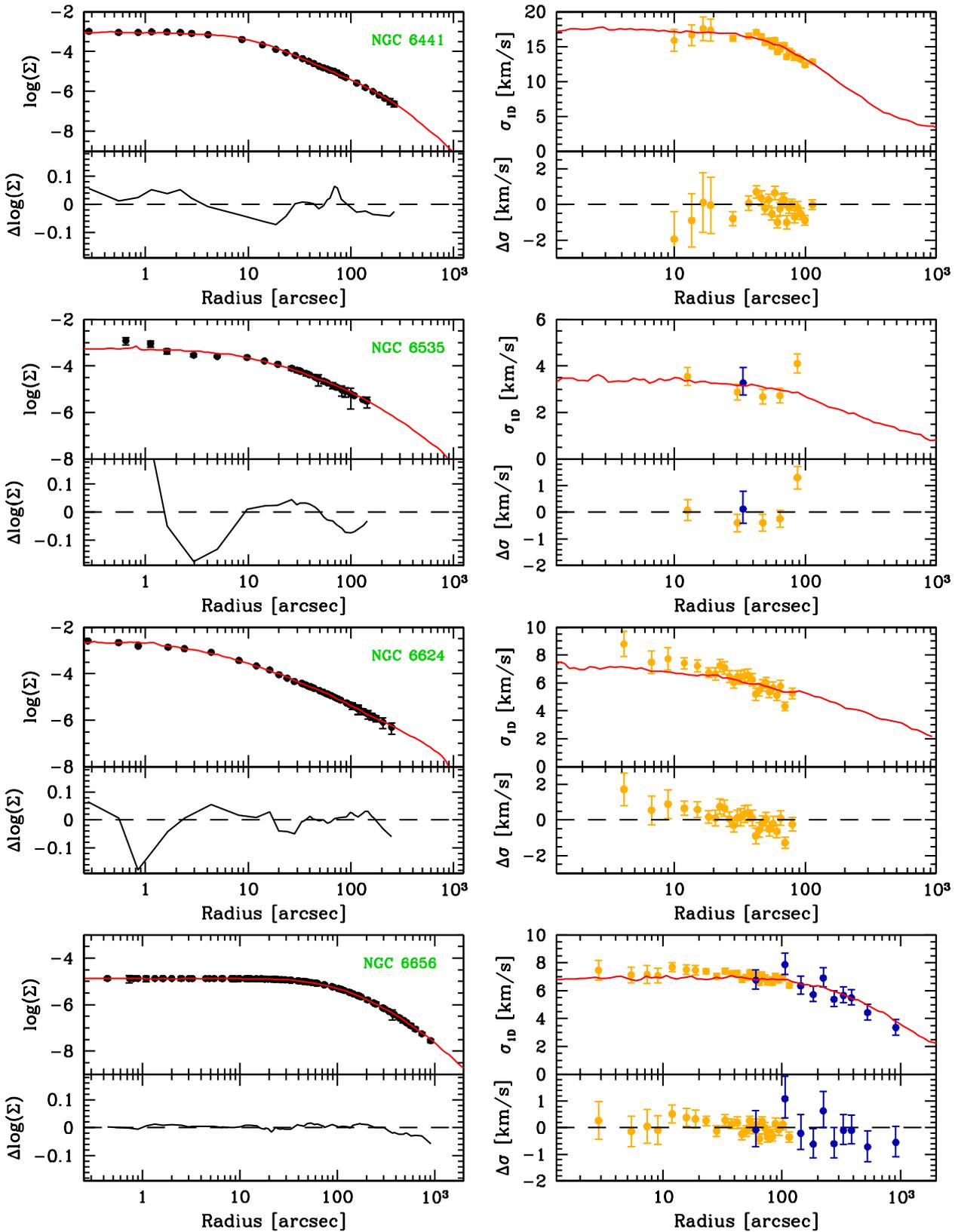}
\end{center}
\caption{Same as Fig. \ref{fig1a} for NGC 6441, NGC 6535, NGC 6624 and NGC 6656.\hspace*{7cm}}
\end{figure*}

\begin{figure*}
\begin{center}
\includegraphics[width=17cm]{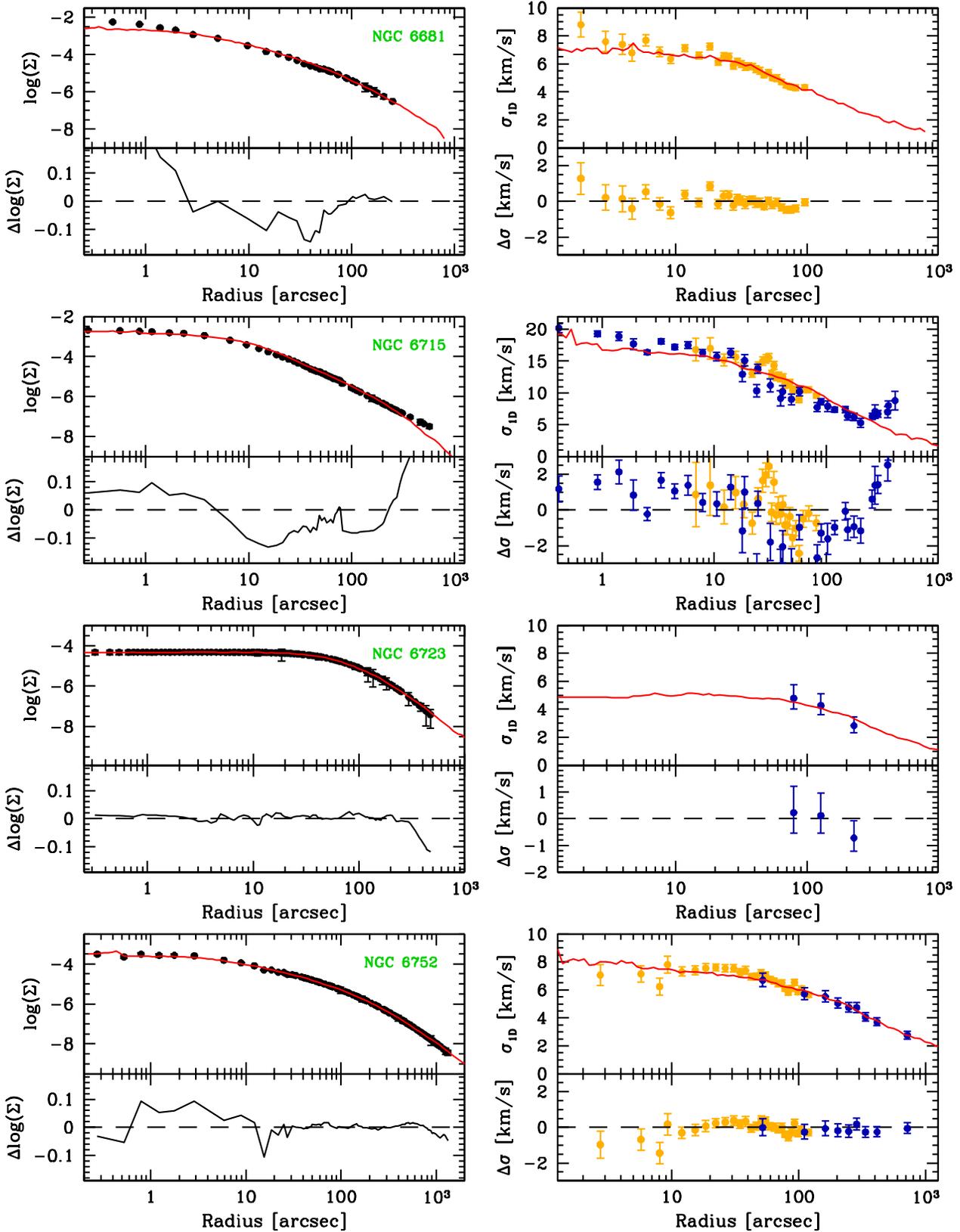}
\end{center}
\caption{Same as Fig. \ref{fig1a} for NGC6681, NGC 6715, NGC 6723 and NGC 6752. The red, solid lines for NGC 6715 show the best-fitting
IMBH model.\hspace*{2cm}}
\end{figure*}

\begin{figure*}
\begin{center}
\includegraphics[width=17cm]{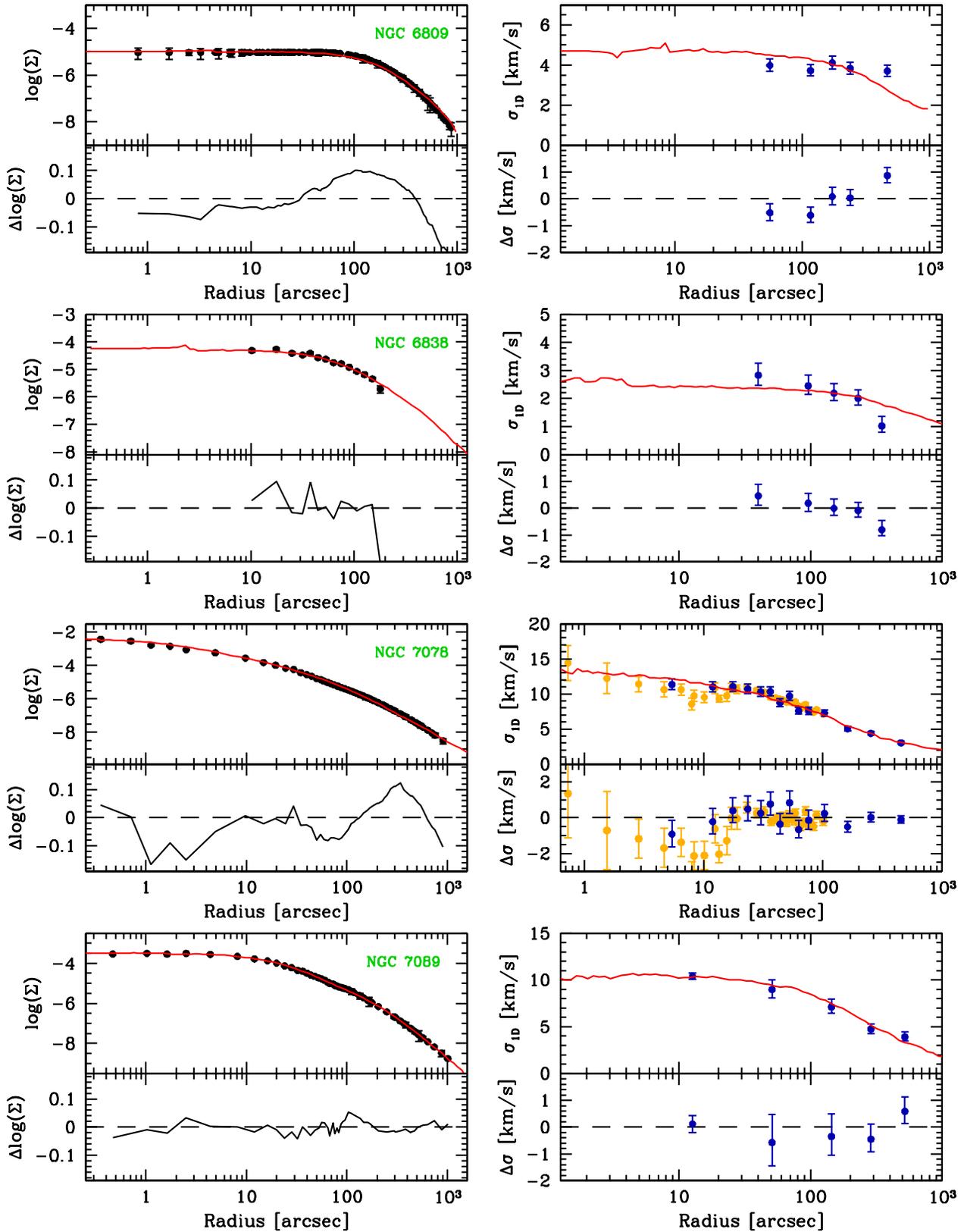}
\end{center}
\caption{Same as Fig. \ref{fig1a} for NGC 6809, NGC 6838, NGC 7078 and NGC 7089.\hspace*{7cm}}
\label{fig19a}
\end{figure*}

\begin{figure*}
\begin{center}
\includegraphics[width=17cm]{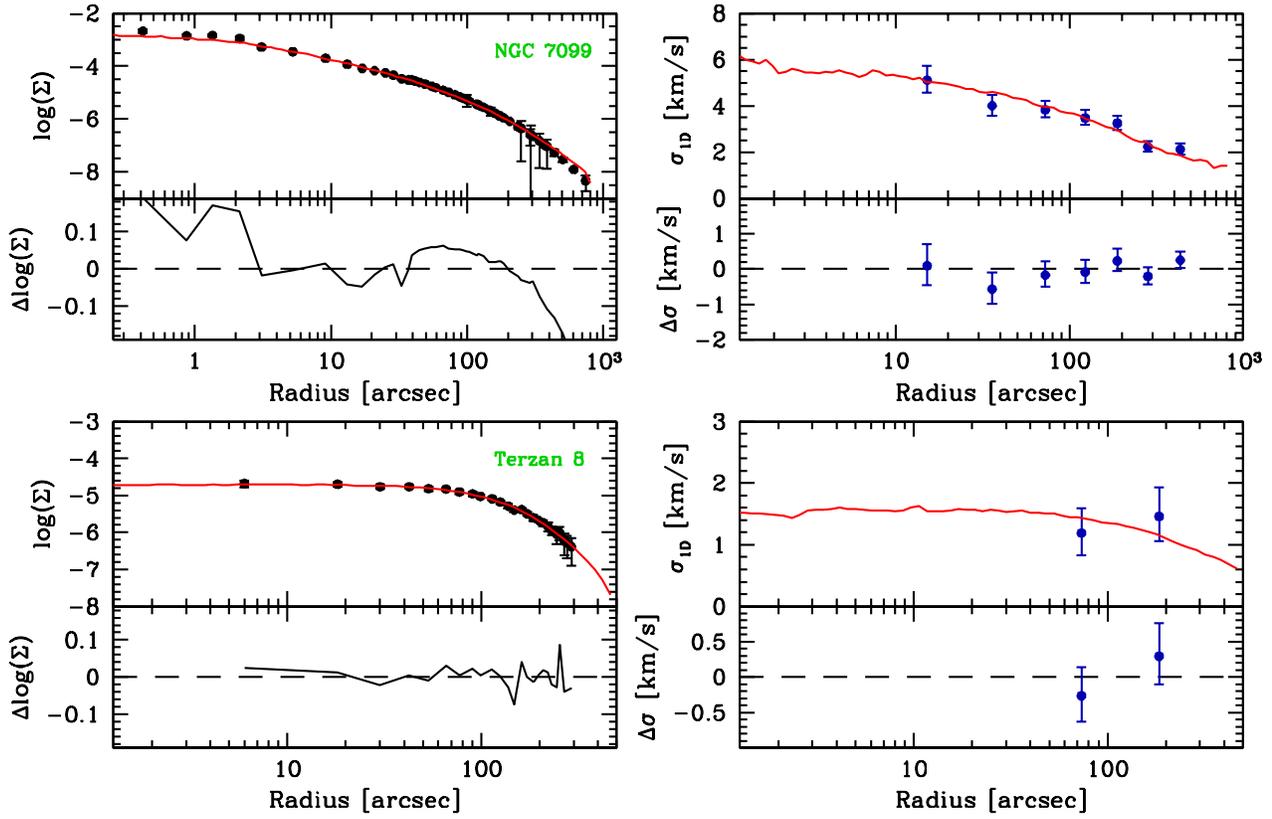}
\end{center}
\caption{Same as Fig. \ref{fig1a} for NGC 7099 and Terzan 8.\hspace*{10.0cm}}
\label{fig13a}
\end{figure*}

\label{lastpage}

\end{document}